\documentclass[prl,twoside,preprintnumbers,superscriptaddress,twocolumn,nofootinbib,nobibnotes]{revtex4}
\usepackage{amsmath,graphicx,slashed,multirow,color,ulem}
\usepackage{caption}
\usepackage{subcaption}
\usepackage{graphicx,epsfig,amssymb,bm,slashed,wasysym,multirow}
\usepackage[utf8]{inputenc}
\def \beq{\begin{equation}}
\def \eeq{\end{equation}}
\def \beqa{\begin{eqnarray}}
\def \eeqa{\end{eqnarray}}

\usepackage{ragged2e}

\usepackage{multirow,booktabs}

\usepackage{graphicx}
\usepackage{url}
\usepackage{cancel}
\usepackage{placeins}
\usepackage{bm}
\usepackage{hyperref}
\usepackage{amsmath}
\usepackage{amssymb}
\usepackage{listings}
\usepackage[toc, page]{appendix}
\usepackage{color}

\usepackage{tabularx}
\newcolumntype{C}[1]{>{\centering\arraybackslash}p{#1}}

\def\bea{\begin{eqnarray}}
\def\eea{\end{eqnarray}}

\def\gsim{\mathrel{\rlap{\lower4pt\hbox{\hskip1pt$\sim$}}
    \raise1pt\hbox{$>$}}}         
\def\lsim{\mathrel{\rlap{\lower4pt\hbox{\hskip1pt$\sim$}}
    \raise1pt\hbox{$<$}}}         

\preprint{DAMTP-2024-19}

\newcommand{\be}{\begin{equation}}
\newcommand{\ee}{\end{equation}}
\newcommand{\bi}{\begin{itemize}}
\newcommand{\ei}{\end{itemize}}
\newcommand{\ben}{\begin{enumerate}}
\newcommand{\een}{\end{enumerate}}

\begin{document}
\title{Unravelling New Physics Signals at the HL-LHC with Low-Energy Constraints}

\author{Elie Hammou}
\email{eh651@cam.ac.uk}
\affiliation{DAMTP, University of Cambridge, Wilberforce Road, Cambridge, CB3 0WA, United Kingdom }

\author{Maria Ubiali }
\email{m.ubiali@damtp.cam.ac.uk }
\affiliation{DAMTP, University of Cambridge, Wilberforce Road, Cambridge, CB3 0WA, United Kingdom }

\date{\today}

\begin{abstract}
Recent studies suggest that global fits of Parton Distribution Functions (PDFs) 
might inadvertently ‘fit away’ signs of new physics in the high-energy tails of the distributions 
measured at the high luminosity programme of the LHC (HL-LHC). 
This could lead to spurious effects that might conceal key BSM signatures and 
hinder the success of indirect searches for new physics.\\
In this letter, we demonstrate that future deep-inelastic scattering (DIS) measurements at the Electron-Ion Collider (EIC), 
and at CERN via FASER$\nu$ and SND@LHC at LHC Run III, and the future neutrino experiments to be hosted at the proposed 
Forward Physics Facility (FPF) at the HL-LHC, provide complementary constraints on large-$x$ sea quarks. 
These constraints are crucial to mitigate the risk of missing key BSM signals, by 
enabling precise constraints on large-$x$ PDFs through a 'BSM-safe' integration of both high- 
and low-energy data, which is essential for a robust interpretation of the high-energy 
measurements.
\end{abstract}

\maketitle


Advancements in experimental analyses and theoretical predictions have ushered the LHC into a new era of precision physics. 
As a proton-proton collider, the accuracy of theoretical predictions heavily depends on the accurate knowledge of 
Parton Distribution Functions (PDFs). 
PDF uncertainties represent a significant portion of the overall theoretical uncertainty 
at the LHC~\cite{Baglio:2022wzu,Amoroso:2022eow,Ubiali:2024pyg}.
Historically, PDFs were primarily extracted from HERA, Tevatron and fixed-target (FT) experiments and 
then used as an input at the LHC. 
However, LHC data now play a pivotal role in global PDF analyses. For instance, 
in the recent NNPDF4.0~\cite{NNPDF:2021njg} determination, 
LHC data constituted nearly a third of the experimental input, providing unique constraints on large-$x$ gluons, quarks, 
and antiquarks. The larger the mass and/or the rapidity of the final state, the larger is the kinematic region in $x$ that 
is probed by the measurement, as at leading order the fraction of the proton's momentum carried by each
parton is given by $x_{1,2}=Q/\sqrt{S} \exp(\pm \eta)$, where $\sqrt{S}$ is the centre-of-mass energy of the experiment, 
$Q$ is the energy scale of the process - roughly of the order of the mass of the final state object(s) - 
and $\eta$ is its rapidity.

While LHC data significantly enhance PDF precision, they are also sensitive to beyond the Standard Model (BSM) 
dynamics. If BSM signals distort an experimental distribution included in a PDF fit—typically assumed to follow the 
SM—this can lead to inconsistencies. These inconsistencies may either exclude the dataset from the fit, ensuring 
consistent PDFs, or be missed by inconsistency tests and cause the PDFs to adapt and incorporate the BSM effects, 
resulting in {\it BSM-biased} PDFs. 
Such considerations cannot be considered purely speculative, as  
it was recently shown~\cite{Hammou:2023heg,Costantini:2024xae} that there exist BSM
scenarios, in particular a scenario involving a new heavy SU(2)$_L$ doublet $W'$ with flavour 
universal coupling $g'$, which would affect the tails of high energy invariant mass Drell-Yan (DY) 
distributions measured at the HL-LHC~\cite{Farina:2016rws,Greljo:2021kvv} 
and which can be completely absorbed, thus fitted away, by a flexible parametrisation of 
the large-$x$ quark and antiquark PDFs.
As a consequence the BSM signals could be missed in indirect new physics searches as the {\it BSM-biased}  
PDFs might adapt in such a way that, when convoluted with the SM predictions, they mimic BSM 
effects. The BSM signal would be lost as it would appear compatible with the SM prediction; schematically, 
%
\begin{equation}
	\hat{\sigma} _{\rm BSM} \otimes f_{\rm SM} \approx \hat{\sigma}_{\rm SM} \otimes f_{\rm BSM},
	\label{eq:PDF_contamination}
\end{equation}
where $\hat{\sigma} _{\rm BSM}$ and $\hat{\sigma} _{\rm SM}$ correspond to the partonic cross sections 
computed according to a given BSM model and to the SM respectively,  
$f_{\rm SM}$ is the {\it true} SM PDF that parametrises the proton's structure, which is obviously decoupled from any 
heavy BSM particles, while $f_{\rm BSM}$ is the {\it BSM-biased} PDF obtained from an inconsistent fit that 
has absorbed the BSM signals in the large-$x$ PDFs.\\

\begin{figure*}[t]
  \begin{center}
    \makebox{\includegraphics[width=\linewidth]{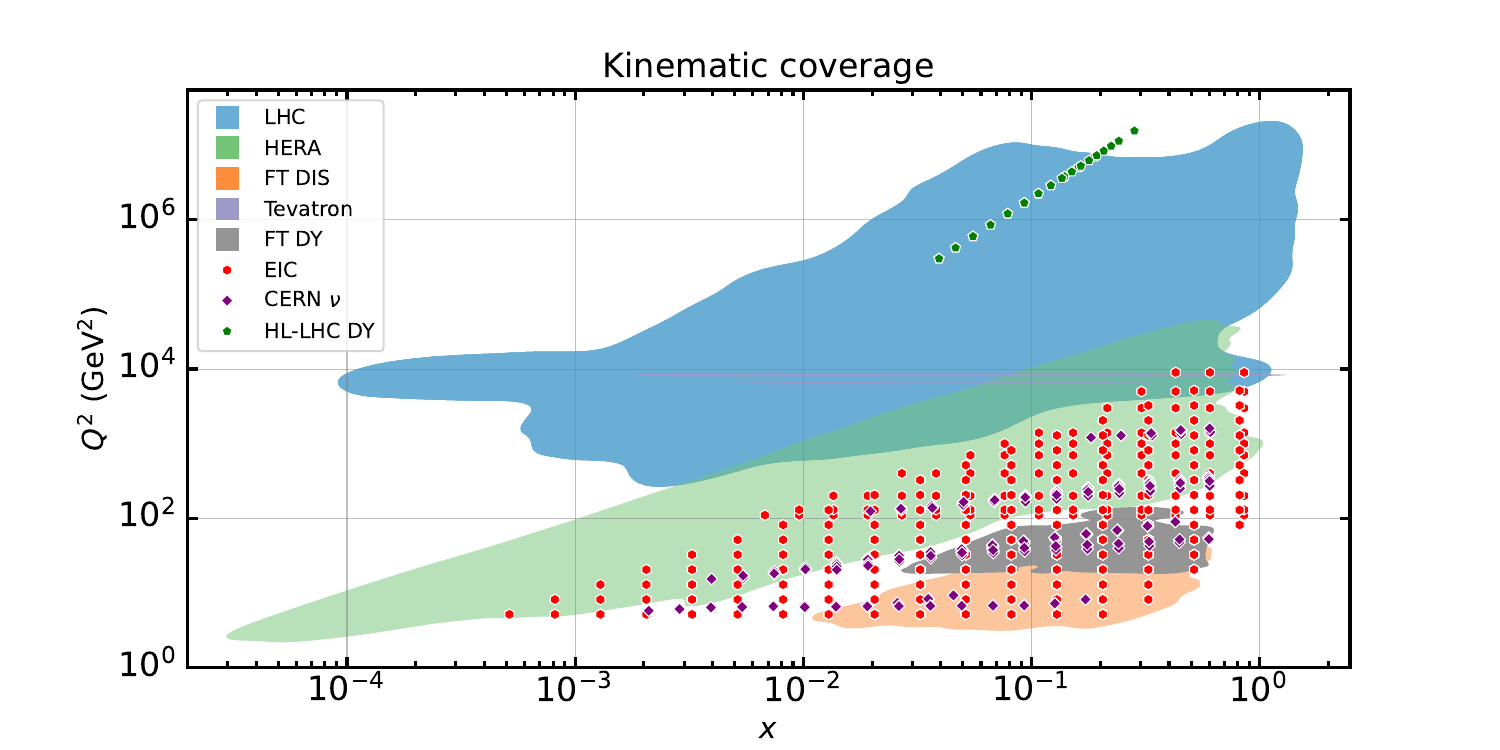}}
   \end{center}
  \vspace{-0.6cm}
  \caption{\small Kinematic coverage of the data (shaded region) and 
  projected data (scattered points) used in this study.}
  \label{fig:kincov}
\end{figure*}

In this letter we show that deep-inelastic scattering (DIS) measurements 
from the Electron Ion Collider (EIC) at Brookhaven National Laboratory~\cite{AbdulKhalek:2021gbh,AbdulKhalek:2022hcn} 
along with neutrino DIS structure functions measurements at CERN from FASER$\nu$~\cite{FASER:2022hcn,FASER:2023zcr,FASER:2024hoe} 
and SND~\cite{SNDLHC:2022ihg,SNDLHC:2023pun} at Run III, 
and at the proposed Forward Physics Facility (FPF) experiments~\cite{Anchordoqui:2021ghd,Feng:2022inv} 
-  FASER$\nu$2, AdvSND, and FLArE - not only provide complementary sensitivity on PDF flavour 
decomposition~\cite{Accardi:2012qut,Aschenauer:2017jsk,Khalek:2021ulf,Cruz-Martinez:2023sdv} 
and on SMEFT Wilson coefficients~\cite{Boughezal:2020uwq,Bissolotti:2023pjh}, but that 
these experiments are essential for indirect BSM searches at high energy.  
The constraints on the large-$x$ sea quarks and gluons that they provide  
are the key for a robust identification of BSM signal at the HL-LHC and at FCC-he and FCC-hh~\cite{Mangano:2022ukr}, 
for example in the Drell-Yan Forward-Backward asymmetry~\cite{Ball:2022qtp,Anataichuk:2023elk,Fiaschi:2021sin}, thus 
proving once more the powerful synergy between diverse experimental programmes. \\

In Fig.~\ref{fig:kincov} we display the kinematic coverage of the data (shaded region) and 
projected data (scattered points) used in this study. The full list is provided in the Appendix.  
The green shaded area corresponds to the HERA measurements~\cite{Aaron:2009aa,Abramowicz:1900rp,Abramowicz:2015mha}, 
the orange area to the Fixed-Target DIS data~\cite{Arneodo:1996qe,Whitlow:1991uw,Benvenuti:1989rh,Onengut:2005kv,Goncharov:2001qe},  
the purple lines to the Tevatron measurements~\cite{Abazov:2008qv,Aaltonen:2011gs,D0:2008hua,Abazov:2008qz,Abazov:2013rja,D0:2014kma}, 
the blue area corresponds to LHC data~\cite{Aad:2011dm,Aaboud:2016btc,Aad:2014qja,Aad:2013iua,Chatrchyan:2012xt,Chatrchyan:2013mza,
Chatrchyan:2013tia,Khachatryan:2016pev,Aaij:2012mda,Aaij:2015gna,LHCb:2015kwa,LHCb:2015mad,ATLAS:2016gic,ATLAS:2017rue,Aaij:2016qqz,
ATLAS:2016fij,Aaij:2016mgv,Aad:2015auj,Khachatryan:2015oaa,ATLAS:2017irc,CMS:2017gbl,Aad:2011fc,Aad:2013lpa,Aad:2014vwa,
Chatrchyan:2012bja,Khachatryan:2015luy,ATLAS:2017kux,CMS:2016lna,Aad:2016xcr,ATLAS:2017nah}
and the grey one to Fixed-Target Drell-Yan (FTDY) measurements~\cite{e665,Towell:2001nh,NuSea:2001idv} included in the fit. 
The scattered points correspond to projected data for the EIC and the CERN neutrino experiments 
as well as for high-energy DY distributions at the HL-LHC.
The main benefit of additional precise measurements covering the relatively low-energy region $Q\lesssim 100$ GeV on top of the 
high-energy region of $Q\approx 1-3$ TeV from the HL-LHC is that such observables, contrarily to the ones measured at the HL-LHC, 
are not susceptible to the effects of the BSM models identified in~\cite{Hammou:2023heg}, and could therefore show a 
tension with the high-energy data, thus flagging a BSM-induced inconsistency in the global PDF fit. 

In this work, as in Ref.~\cite{Hammou:2023heg}, results are based on closure-test fits, which allow 
a systematic study of BSM-induced effects in PDF fits by isolating them from all other effects such as 
missing higher order uncertainties in theory predictions and experimental inconsistencies. 
Closure-test fits are done by generating the central values of the data included in the fit according to a known underlying law, 
given by a model and a perturbative order that determine the {\it true} partonic cross sections and a reference set of 
PDFs representing the {\it true} structure of the proton, and let them fluctuate according to experimental uncertainties. 
If a PDF fitting methodology is successful and the model used in the fit is the same as the model used in the data generation, 
then the fitted PDFs should reproduce the underlying law and the $\chi^2$ per datapoint should 
be centred around one for all experiments. If however the data are generated according to a given BSM model but PDFs 
are fitted assuming the SM, then a closure test is successful only if the PDFs manage to {\it absorb} the BSM shifts, as 
schematically indicated in Eq.~\eqref{eq:PDF_contamination}. More details about closure tests and 
related statistical estimators to estimate inconsistencies are given in the Appendix. 
\begin{figure}[b!]
  \begin{center}
    \makebox{\includegraphics[width=0.9\columnwidth]{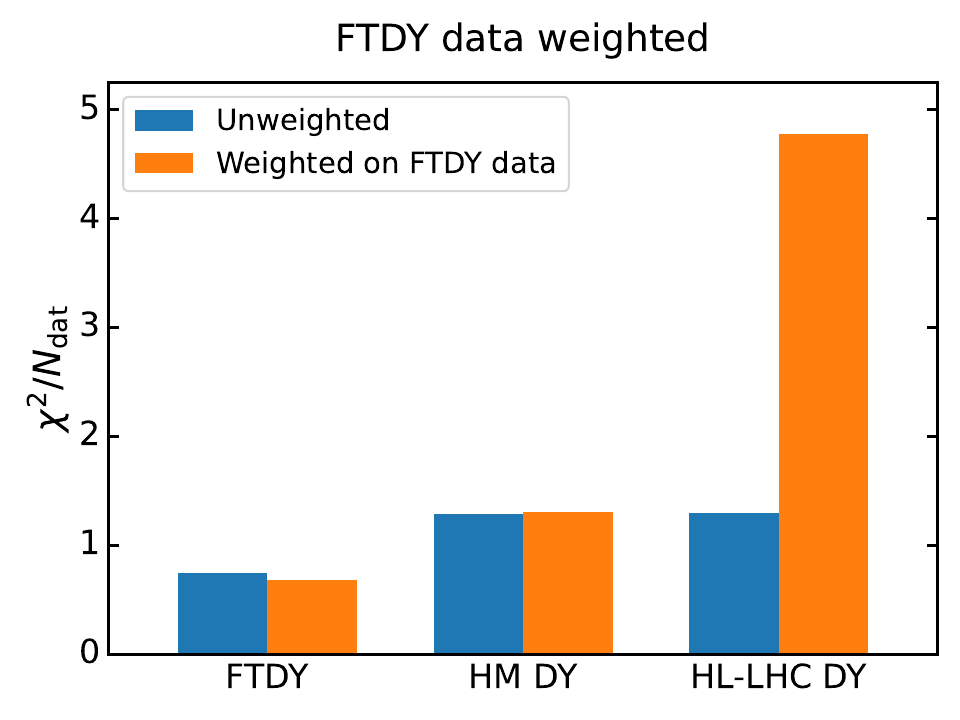}}
   \end{center}
  \vspace{-0.6cm}
  \caption{\small Data-theory agreement for fixed-target DY data (FTDY), LHC Run I and Run II 
  high-mass DY data (HM DY) and HL-LHC high-mass DY projections (HL-LHC) in two closure-test fits. 
  In both fits the BSM model described in the main text is injected in the data and PDFs are fitted 
  assuming the SM. The height of the blue histogram bars (unweighted) represents the $\chi^2$ per 
  datapoint obtained in a global NNPDF4.0-like closure-test fit augmented by the HL-LHC projections. 
  The height of the orange bars (weighted on FTDY data) corresponds to the same quantity 
  obtained in a similar closure test, in which more weight is given to the FTDY data. }
  \label{fig:FTDY_weighted}
\end{figure}

In Fig.~\ref{fig:FTDY_weighted} we compare the data-theory agreement for three subsets of the data included in 
two closure-test fits. 
In both fits, the data is generated by injecting the $W'$ model discussed in~\cite{Farina:2016rws,Greljo:2021kvv,Hammou:2023heg} 
with $M_{W'}=13.8$ TeV and $g'=1$, and PDFs are fitted assuming the SM. Details of the model under consideration are summarised 
in the Appendix. 
The three experimental subsets correspond to fixed-target Drell-Yan (FTDY) data, the full LHC Run I and Run II high-mass DY 
data (HM DY) included in the NNPDF4.0 global fit~\cite{Aad:2013iua,Chatrchyan:2013tia,Aad:2016zzw,CMS:2014jea,CMS:2018mdl}, 
and HL-LHC high-mass DY projections (HL-LHC) that were considered in~\cite{Khalek:2018mdn,Greljo:2021kvv,Iranipour:2022iak,Hammou:2023heg,Costantini:2024xae} 
to assess the potential impact of high-mass HL-LHC DY distributions on the PDFs. 
The height of the histogram bars represents the $\chi^2$ per datapoint for 
each data category as obtained in two fits: a closure test performed on the global NNPDF4.0 dataset augmented by 
the HL-LHC projections (blue bars) and the same closure test in which more weight is given to the FTDY data (orange bars). 
Details about how weighted fits are performed are given in the Appendix. The BSM model that we inject in the data affects 
mostly the  HL-LHC DY projections and, although moderately, the Run I and Run II high-mass DY 
distributions. 
As we can see, the closure-test fit associated to the blue bars is successful, indicating that the 
PDF parametrisation absorbs the effects of the BSM model injected in the data. In particular 
the fit quality of the HL-LHC data is extremely good, thus the BSM-induced inconsistency would not be 
spotted in the consistency checks across datasets that PDF fitters routinely run.  
The orange bars instead correspond to a fit in which the FTDY data was given more weight, thus features smaller uncertainties and 
has a stronger pull on the fit. In that case the $\chi ^2$ per data point for the FTDY data 
improves, while the one of the HL-LHC projections deteriorates to a level that an inconsistency would 
be spotted and these distributions would not be included in a global PDF analysis. 
Typically in a PDF fit an experiment consisting of $n_{\rm dat}$ datapoints is considered inconsistent with the bulk of the other 
data included in the fit if the $\chi^2$ per datapoint exceeds a critical threshold of $\chi^2_{\rm thr}=1.5$, and 
if its deviation from 1 in terms of the standard deviation of a $\chi^2$ distribution exceeds the critical threshold of 2, 
namely $n_{\sigma}=(\chi^2-n_{\rm dat})/\sqrt{2 n_{\rm dat}}>2$. 
To summarise, in the standard fit (blue bars) the FTDY data is not precise enough to flag the BSM-induced bias in the PDFs. 
However, when extra weight is given to it, the $\chi ^2$ of the HL-LHC increases dramatically, flagging the BSM-induced bias (orange bars). 
The reason behind this finding is that the low-energy FTDY data, thanks to a combination of proton and isoscalar targets 
constrain the $\bar{u}/\bar{d}$ ratio at large $x$, thus increasing the tension with the HL-LHC pull on the anti-quarks.\\

While FTDY experiments are not currently planned by the HEP community, the 
EIC~\cite{Aschenauer:2017jsk,Accardi:2012qut} has been 
approved by the United States Department of Energy at Brookhaven National Laboratory, and
could record the first scattering events as early as 2030.

Our analysis starts by considering the constraining power of the EIC projections that were 
considered in~\cite{Khalek:2021ulf}. The datapoints added to the PDF fits are listed in 
Table~\ref{tab:EIC_data} in the Appendix, and their kinematic coverage is 
displayed in Fig.~\ref{fig:kincov}. We consider measurements of charged-current (CC) and 
neutral-current (NC) DIS processes, using both electrons and positrons as projectiles and 
both protons and neutrons as targets. 
Although the EIC will be able to use a greater variety of heavy ions as targets, 
we focus on the ones susceptible to constrain the proton PDF. We added these observables 
to the PDF fit dataset alongside the projected data from the HL-LHC that is susceptible to 
signs of the BSM model considered in this work and performed several fits for different $W'$ 
masses by keeping its flavour-universal coupling $g'=1$. 

\begin{figure}[tb!]
  \begin{center}
    \makebox{\includegraphics[width=\columnwidth]{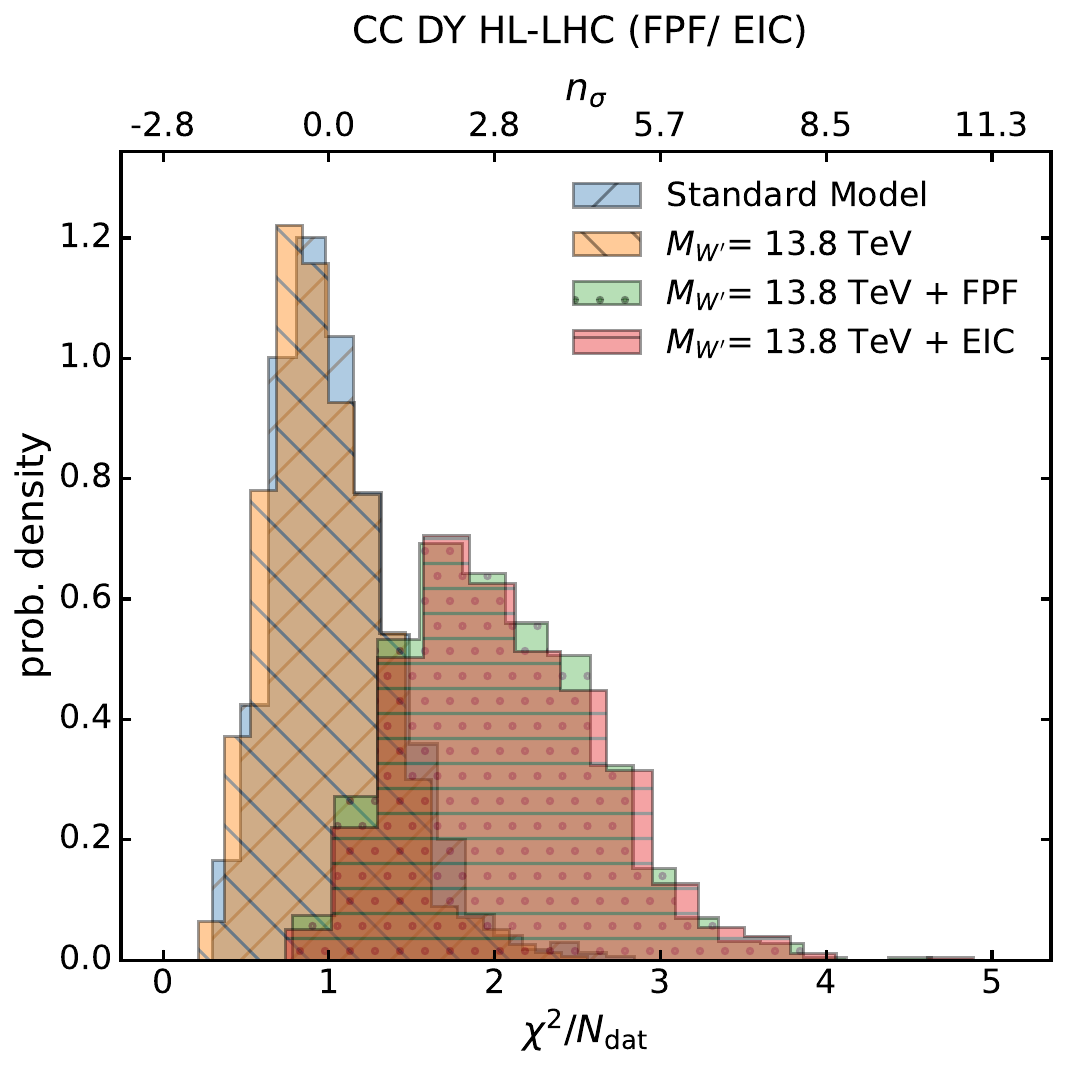}}
  \end{center}
  \vspace{-0.5cm}
  \caption{\small Comparison of the distribution of the $\chi^2$ per datapoint (bottom $x$-axis) and 
  $n_\sigma$ (top $x$-axis) for the high-mass HL-LHC DY CC projected data between
  a consistent fit in which both data and predictions are generated according to the SM (blue histogram), 
  the fit biased by BSM including the HL-LHC projected data (orange histogram) and the 
  same fit including both the HL-LHC and the EIC (red histogram) or the FPF (green histogram) projected 
  data respectively.}
  \label{fig:chi2_EIC_FPF}
\end{figure}

The main observation is that the inclusion of EIC projections alongside the HL-LHC ones makes 
a significant difference in the ability of the large-$x$ antiquark distributions to shift and 
fit away the BSM signal injected in the tails of the DY distributions. This can be observed in 
Fig.~\ref{fig:chi2_EIC_FPF}. There we display the $\chi^{2(k)}$ and $n_\sigma^{(k)}$ distributions 
across 1000 random instances $(k)$ of the data generation in the closure-test fits for the datasets 
that are mostly affected by BSM corrections, in this case the HL-LHC CC DY projections. 
Looking at the blue and orange histograms, we observe that  
the global closure-test fit (in which data are generated according to a $W'$ model with $M_{W'} = 13.8$ TeV 
and PDFs are fitted assuming the SM) displays the same good fit quality 
as the baseline (in which data are generated according to the SM and PDFs are fitted assuming the SM). 
As it was pointed out in Ref.~\cite{Hammou:2023heg} the value of $M_{W'}=13.8$ TeV corresponds to the 
{\it maximal BSM threshold}, {\it i.~e.} to the lowest $M_{W'}$ that can be absorbed by the PDFs 
without deteriorating the $\chi^2$ of the fit, particularly of the BSM-affected data. 
The situation changes if the EIC projections as included. Indeed the red dotted histogram shows 
that the inclusion of this data would prevent PDFs from absorbing new physics, 
as the peak of the $\chi^{2}$ and $n_{\sigma}$ distributions shifts beyond the critical values 
of 1.5 and 2.0 respectively that would flag up the HL-LHC CC DY projections as inconsistent with the 
bulk of the datasets included in the analysis, hence it would exclude it from the global PDF fit.
%

Next we assess the effect of the CERN neutrino data. The recent observation of LHC neutrinos 
by the FASER$\nu$~\cite{FASER:2022hcn,FASER:2023zcr,FASER:2024hoe} and 
SND~\cite{SNDLHC:2022ihg,SNDLHC:2023pun} far-forward experiments demonstrates that neutrino beams produced by $pp$ 
collisions at the LHC can be deployed for physics studies~\cite{Cruz-Martinez:2023sdv}. 
Beyond the ongoing Run III, a dedicated suite of upgraded far-forward neutrino experiments would
be hosted by the proposed Forward Physics Facility (FPF)~\cite{Anchordoqui:2021ghd,Feng:2022inv} 
operating concurrently with the HL-LHC.
We now investigate whether the projected measurements at the 
two facilities considered in~\cite{Cruz-Martinez:2023sdv} might provide a similar handle on the 
proton structure as the EIC structure function measurements, which would prevent BSM-induced bias in 
PDF fits. 
Here we only consider CC DIS processes, where the projectiles are neutrinos
and anti-neutrinos, coming from the LHC beam, hitting isoscalar targets and the processes associated 
with charm production. As in~\cite{Cruz-Martinez:2023sdv} we ignore the SND Run III projections 
as the current statistics is too little ro produce faithful projected data. 
The details of the datasets are presented in Tab.~\ref{tab:FPF_data} in the Appendix 
and the kinematic coverage is displayed in Fig.~\ref{fig:kincov}.
Looking at the green histogram in Fig.~\ref{fig:chi2_EIC_FPF}, we observe that, 
analogously to what happens in the case of the EIC, the inclusion of the FPF projections 
would shift the $\chi^2$ distribution for the HL-LHC high mass DY data and would flag them  
up as inconsistent with the bulk of other data included in the fit. 
Note that the main constraint comes from the proposed upgraded far-forward neutrino experiments as 
FASER$\nu$ has little impact on the PDFs. 

\begin{figure}[tb!]
  \begin{center}
    \makebox{\includegraphics[width=\columnwidth]{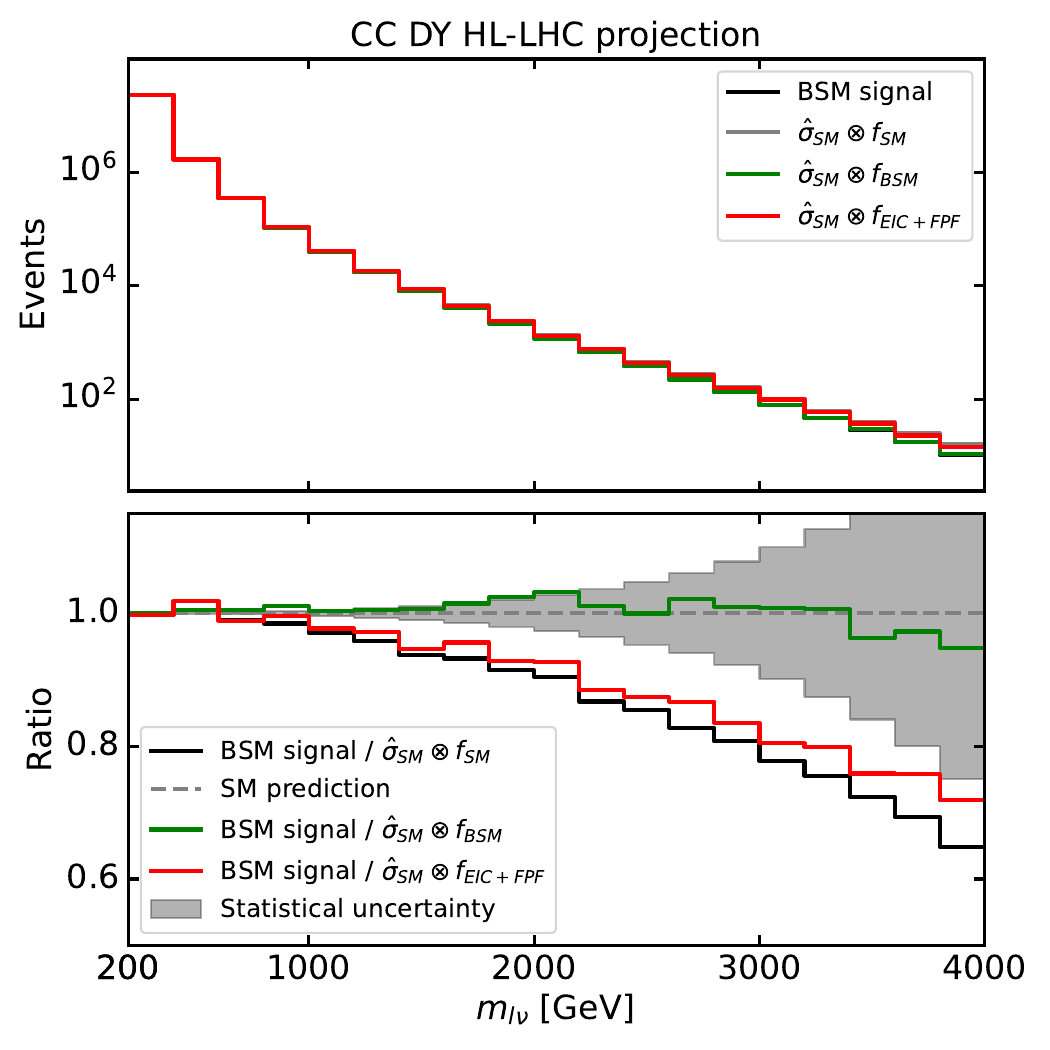}}
  \end{center}
  \vspace{-0.5cm}
  \caption{\small Predictions for CC DY ($pp\to l^-\bar{\nu}_l$) inclusive cross sections
  differential in $M_{l\nu}$, with $l= (e,\mu)$. Factorisation and renormalisation scales are set to
  $\mu_F=\mu_R=M_T$, where $M_T$ is the transverse mass of each bin. The BSM signal is compared to 
  the SM predictions obtained with a {\it BSM-biased} set of 
  PDFs (green line) or with a set of PDFs obtained from a joint fit of the HL-LHC data and the 
  FPF+EIC data (red line), 
  or finally with the {\it true} SM PDFs (black line). In the bottom inset the predictions obtained 
  with the various PDF sets are compared to the SM prediction and its total uncertainty.}
  \label{fig:histo}
\end{figure}

The main finding is summarised in Fig.~\ref{fig:histo}, in which we compare the data, 
featuring the BSM signal, {\it i.~e.} the depletion in the tails associated with a new heavy $W'$ 
with $M_{W'}=13.8$ TeV and $g'=1$, to the SM theoretical predictions obtained with a {\it BSM-biased}  
set of PDFs (green line) or with a set of PDFs obtained from a joint fit of the HL-LHC data and 
the FPF+EIC data, in which the CC HL-LHC data get flagged and excluded from the fit 
(red line), or finally with the {\it true} SM PDFs (black line).
The inset shows the ratio between the BSM signal and the SM predictions for each 
of the PDF sets. 

The green line closely matching the SM prediction demonstrates that the {\it BSM-biased} PDFs mimic the BSM signals present in the data
and erase the deviation from the SM that should be observed, as in Eq.~\eqref{eq:PDF_contamination}. 
As a result, using these PDFs in theoretical predictions would hide the BSM signal. 
This wouldn't occur if the {\it true} PDFs (represented by the black line, which we wouldn't know in reality) were used. 
However, the red line—derived from fitting the global dataset, including only the NC HL-LHC data compatible with EIC+FPF 
data—closely matches the black line and would allow us to detect the BSM signal.

To summarise, in this work we have identified low-energy datasets that constrain the large-$x$ region and 
can be included in PDF fits, enabling the safe integration of high-energy constraints from the HL-LHC. 
Our results show that potential BSM effects, which might otherwise be absorbed into the PDFs, 
can be fully disentangled in high-energy measurements. Incorporating EIC and FPF measurements in a 
global PDF analysis will help minimise the risk of BSM-induced bias in PDF fits and allow for more 
consistent identification of BSM effects in high-energy data, which can then be analysed separately.

A natural follow-up of our current study would be the extension to more complex BSM scenarios affecting both the 
large-$x$ quarks and gluons. Both the code and the projected data used in this study are publicly available 
at the {\tt SIMUnet}~\cite{Costantini:2024xae} repository\footnote{https://github.com/HEP-PBSP/SIMUnet}.
Users are encouraged to test the robustness of our findings against different new physics scenarios and experimental 
projections.

\paragraph{Acknowledgements.}
We are grateful to Juan Rojo for his precious suggestions concerning the presentation of our results. 
We thank Juan Rojo, Tanjona Rabemananjara and Juan Cruz-Martinez for providing us with the FPF projections and 
the corresponding theoretical 
predictions. We thank Emanuele Nocera for his assistance in identifying the EIC projected data and in 
generating theoretical predictions. We are most grateful for Felix Hekhorn for his precious help in using 
EKO and YADISM to generate interpolating grids for DIS observables. 
E.~H. and M.U. are supported by the
European Research Council under the European Union’s Horizon 2020
research and innovation Programme (grant agreement n.950246), and partially by the STFC consolidated grant
ST/T000694/1 and ST/X000664/1.

\bibliography{letter.bib}

\begin{thebibliography}{82}
\expandafter\ifx\csname natexlab\endcsname\relax\def\natexlab#1{#1}\fi
\expandafter\ifx\csname bibnamefont\endcsname\relax
  \def\bibnamefont#1{#1}\fi
\expandafter\ifx\csname bibfnamefont\endcsname\relax
  \def\bibfnamefont#1{#1}\fi
\expandafter\ifx\csname citenamefont\endcsname\relax
  \def\citenamefont#1{#1}\fi
\expandafter\ifx\csname url\endcsname\relax
  \def\url#1{\texttt{#1}}\fi
\expandafter\ifx\csname urlprefix\endcsname\relax\def\urlprefix{URL }\fi
\providecommand{\bibinfo}[2]{#2}
\providecommand{\eprint}[2][]{\url{#2}}

\bibitem[{\citenamefont{Baglio et~al.}(2022)\citenamefont{Baglio, Duhr,
  Mistlberger, and Szafron}}]{Baglio:2022wzu}
\bibinfo{author}{\bibfnamefont{J.}~\bibnamefont{Baglio}},
  \bibinfo{author}{\bibfnamefont{C.}~\bibnamefont{Duhr}},
  \bibinfo{author}{\bibfnamefont{B.}~\bibnamefont{Mistlberger}},
  \bibnamefont{and} \bibinfo{author}{\bibfnamefont{R.}~\bibnamefont{Szafron}},
  \bibinfo{journal}{JHEP} \textbf{\bibinfo{volume}{12}}, \bibinfo{pages}{066}
  (\bibinfo{year}{2022}), \eprint{2209.06138}.

\bibitem[{\citenamefont{Amoroso et~al.}(2022)}]{Amoroso:2022eow}
\bibinfo{author}{\bibfnamefont{S.}~\bibnamefont{Amoroso}} \bibnamefont{et~al.},
  \bibinfo{journal}{Acta Phys. Polon. B} \textbf{\bibinfo{volume}{53}},
  \bibinfo{pages}{12} (\bibinfo{year}{2022}), \eprint{2203.13923}.

\bibitem[{\citenamefont{Ubiali}(2024)}]{Ubiali:2024pyg}
\bibinfo{author}{\bibfnamefont{M.}~\bibnamefont{Ubiali}}
  (\bibinfo{year}{2024}), \eprint{2404.08508}.

\bibitem[{\citenamefont{Ball et~al.}(2022{\natexlab{a}})}]{NNPDF:2021njg}
\bibinfo{author}{\bibfnamefont{R.~D.} \bibnamefont{Ball}} \bibnamefont{et~al.}
  (\bibinfo{collaboration}{NNPDF}), \bibinfo{journal}{Eur. Phys. J. C}
  \textbf{\bibinfo{volume}{82}}, \bibinfo{pages}{428}
  (\bibinfo{year}{2022}{\natexlab{a}}), \eprint{2109.02653}.

\bibitem[{\citenamefont{Hammou et~al.}(2023)\citenamefont{Hammou, Kassabov,
  Madigan, Mangano, Mantani, Moore, Alvarado, and Ubiali}}]{Hammou:2023heg}
\bibinfo{author}{\bibfnamefont{E.}~\bibnamefont{Hammou}},
  \bibinfo{author}{\bibfnamefont{Z.}~\bibnamefont{Kassabov}},
  \bibinfo{author}{\bibfnamefont{M.}~\bibnamefont{Madigan}},
  \bibinfo{author}{\bibfnamefont{M.~L.} \bibnamefont{Mangano}},
  \bibinfo{author}{\bibfnamefont{L.}~\bibnamefont{Mantani}},
  \bibinfo{author}{\bibfnamefont{J.}~\bibnamefont{Moore}},
  \bibinfo{author}{\bibfnamefont{M.~M.} \bibnamefont{Alvarado}},
  \bibnamefont{and} \bibinfo{author}{\bibfnamefont{M.}~\bibnamefont{Ubiali}},
  \bibinfo{journal}{JHEP} \textbf{\bibinfo{volume}{11}}, \bibinfo{pages}{090}
  (\bibinfo{year}{2023}), \eprint{2307.10370}.

\bibitem[{\citenamefont{Costantini et~al.}(2024)\citenamefont{Costantini,
  Hammou, Kassabov, Madigan, Mantani, Morales~Alvarado, Moore, and
  Ubiali}}]{Costantini:2024xae}
\bibinfo{author}{\bibfnamefont{M.~N.} \bibnamefont{Costantini}},
  \bibinfo{author}{\bibfnamefont{E.}~\bibnamefont{Hammou}},
  \bibinfo{author}{\bibfnamefont{Z.}~\bibnamefont{Kassabov}},
  \bibinfo{author}{\bibfnamefont{M.}~\bibnamefont{Madigan}},
  \bibinfo{author}{\bibfnamefont{L.}~\bibnamefont{Mantani}},
  \bibinfo{author}{\bibfnamefont{M.}~\bibnamefont{Morales~Alvarado}},
  \bibinfo{author}{\bibfnamefont{J.~M.} \bibnamefont{Moore}}, \bibnamefont{and}
  \bibinfo{author}{\bibfnamefont{M.}~\bibnamefont{Ubiali}}
  (\bibinfo{collaboration}{PBSP}), \bibinfo{journal}{Eur. Phys. J. C}
  \textbf{\bibinfo{volume}{84}}, \bibinfo{pages}{805} (\bibinfo{year}{2024}),
  \eprint{2402.03308}.

\bibitem[{\citenamefont{Farina et~al.}(2017)\citenamefont{Farina, Panico,
  Pappadopulo, Ruderman, Torre, and Wulzer}}]{Farina:2016rws}
\bibinfo{author}{\bibfnamefont{M.}~\bibnamefont{Farina}},
  \bibinfo{author}{\bibfnamefont{G.}~\bibnamefont{Panico}},
  \bibinfo{author}{\bibfnamefont{D.}~\bibnamefont{Pappadopulo}},
  \bibinfo{author}{\bibfnamefont{J.~T.} \bibnamefont{Ruderman}},
  \bibinfo{author}{\bibfnamefont{R.}~\bibnamefont{Torre}}, \bibnamefont{and}
  \bibinfo{author}{\bibfnamefont{A.}~\bibnamefont{Wulzer}},
  \bibinfo{journal}{Phys. Lett.} \textbf{\bibinfo{volume}{B772}},
  \bibinfo{pages}{210} (\bibinfo{year}{2017}), \eprint{1609.08157}.

\bibitem[{\citenamefont{Greljo et~al.}(2021)\citenamefont{Greljo, Iranipour,
  Kassabov, Madigan, Moore, Rojo, Ubiali, and Voisey}}]{Greljo:2021kvv}
\bibinfo{author}{\bibfnamefont{A.}~\bibnamefont{Greljo}},
  \bibinfo{author}{\bibfnamefont{S.}~\bibnamefont{Iranipour}},
  \bibinfo{author}{\bibfnamefont{Z.}~\bibnamefont{Kassabov}},
  \bibinfo{author}{\bibfnamefont{M.}~\bibnamefont{Madigan}},
  \bibinfo{author}{\bibfnamefont{J.}~\bibnamefont{Moore}},
  \bibinfo{author}{\bibfnamefont{J.}~\bibnamefont{Rojo}},
  \bibinfo{author}{\bibfnamefont{M.}~\bibnamefont{Ubiali}}, \bibnamefont{and}
  \bibinfo{author}{\bibfnamefont{C.}~\bibnamefont{Voisey}},
  \bibinfo{journal}{JHEP} \textbf{\bibinfo{volume}{07}}, \bibinfo{pages}{122}
  (\bibinfo{year}{2021}), \eprint{2104.02723}.

\bibitem[{\citenamefont{Abdul~Khalek
  et~al.}(2022{\natexlab{a}})}]{AbdulKhalek:2021gbh}
\bibinfo{author}{\bibfnamefont{R.}~\bibnamefont{Abdul~Khalek}}
  \bibnamefont{et~al.}, \bibinfo{journal}{Nucl. Phys. A}
  \textbf{\bibinfo{volume}{1026}}, \bibinfo{pages}{122447}
  (\bibinfo{year}{2022}{\natexlab{a}}), \eprint{2103.05419}.

\bibitem[{\citenamefont{Abdul~Khalek
  et~al.}(2022{\natexlab{b}})}]{AbdulKhalek:2022hcn}
\bibinfo{author}{\bibfnamefont{R.}~\bibnamefont{Abdul~Khalek}}
  \bibnamefont{et~al.} (\bibinfo{year}{2022}{\natexlab{b}}),
  \eprint{2203.13199}.

\bibitem[{\citenamefont{Abreu et~al.}(2024)}]{FASER:2022hcn}
\bibinfo{author}{\bibfnamefont{H.}~\bibnamefont{Abreu}} \bibnamefont{et~al.}
  (\bibinfo{collaboration}{FASER}), \bibinfo{journal}{JINST}
  \textbf{\bibinfo{volume}{19}}, \bibinfo{pages}{P05066}
  (\bibinfo{year}{2024}), \eprint{2207.11427}.

\bibitem[{\citenamefont{Abreu et~al.}(2023)}]{FASER:2023zcr}
\bibinfo{author}{\bibfnamefont{H.}~\bibnamefont{Abreu}} \bibnamefont{et~al.}
  (\bibinfo{collaboration}{FASER}), \bibinfo{journal}{Phys. Rev. Lett.}
  \textbf{\bibinfo{volume}{131}}, \bibinfo{pages}{031801}
  (\bibinfo{year}{2023}), \eprint{2303.14185}.

\bibitem[{\citenamefont{Mammen~Abraham et~al.}(2024)}]{FASER:2024hoe}
\bibinfo{author}{\bibfnamefont{R.}~\bibnamefont{Mammen~Abraham}}
  \bibnamefont{et~al.} (\bibinfo{collaboration}{FASER}),
  \bibinfo{journal}{Phys. Rev. Lett.} \textbf{\bibinfo{volume}{133}},
  \bibinfo{pages}{021802} (\bibinfo{year}{2024}), \eprint{2403.12520}.

\bibitem[{\citenamefont{Acampora et~al.}(2024)}]{SNDLHC:2022ihg}
\bibinfo{author}{\bibfnamefont{G.}~\bibnamefont{Acampora}} \bibnamefont{et~al.}
  (\bibinfo{collaboration}{SND@LHC}), \bibinfo{journal}{JINST}
  \textbf{\bibinfo{volume}{19}}, \bibinfo{pages}{P05067}
  (\bibinfo{year}{2024}), \eprint{2210.02784}.

\bibitem[{\citenamefont{Albanese et~al.}(2023)}]{SNDLHC:2023pun}
\bibinfo{author}{\bibfnamefont{R.}~\bibnamefont{Albanese}} \bibnamefont{et~al.}
  (\bibinfo{collaboration}{SND@LHC}), \bibinfo{journal}{Phys. Rev. Lett.}
  \textbf{\bibinfo{volume}{131}}, \bibinfo{pages}{031802}
  (\bibinfo{year}{2023}), \eprint{2305.09383}.

\bibitem[{\citenamefont{Anchordoqui et~al.}(2022)}]{Anchordoqui:2021ghd}
\bibinfo{author}{\bibfnamefont{L.~A.} \bibnamefont{Anchordoqui}}
  \bibnamefont{et~al.}, \bibinfo{journal}{Phys. Rept.}
  \textbf{\bibinfo{volume}{968}}, \bibinfo{pages}{1} (\bibinfo{year}{2022}),
  \eprint{2109.10905}.

\bibitem[{\citenamefont{Feng et~al.}(2023)}]{Feng:2022inv}
\bibinfo{author}{\bibfnamefont{J.~L.} \bibnamefont{Feng}} \bibnamefont{et~al.},
  \bibinfo{journal}{J. Phys. G} \textbf{\bibinfo{volume}{50}},
  \bibinfo{pages}{030501} (\bibinfo{year}{2023}), \eprint{2203.05090}.

\bibitem[{\citenamefont{Accardi et~al.}(2016)}]{Accardi:2012qut}
\bibinfo{author}{\bibfnamefont{A.}~\bibnamefont{Accardi}} \bibnamefont{et~al.},
  \bibinfo{journal}{Eur. Phys. J. A} \textbf{\bibinfo{volume}{52}},
  \bibinfo{pages}{268} (\bibinfo{year}{2016}), \eprint{1212.1701}.

\bibitem[{\citenamefont{Aschenauer et~al.}(2019)\citenamefont{Aschenauer,
  Fazio, Lee, Mantysaari, Page, Schenke, Ullrich, Venugopalan, and
  Zurita}}]{Aschenauer:2017jsk}
\bibinfo{author}{\bibfnamefont{E.~C.} \bibnamefont{Aschenauer}},
  \bibinfo{author}{\bibfnamefont{S.}~\bibnamefont{Fazio}},
  \bibinfo{author}{\bibfnamefont{J.~H.} \bibnamefont{Lee}},
  \bibinfo{author}{\bibfnamefont{H.}~\bibnamefont{Mantysaari}},
  \bibinfo{author}{\bibfnamefont{B.~S.} \bibnamefont{Page}},
  \bibinfo{author}{\bibfnamefont{B.}~\bibnamefont{Schenke}},
  \bibinfo{author}{\bibfnamefont{T.}~\bibnamefont{Ullrich}},
  \bibinfo{author}{\bibfnamefont{R.}~\bibnamefont{Venugopalan}},
  \bibnamefont{and} \bibinfo{author}{\bibfnamefont{P.}~\bibnamefont{Zurita}},
  \bibinfo{journal}{Rept. Prog. Phys.} \textbf{\bibinfo{volume}{82}},
  \bibinfo{pages}{024301} (\bibinfo{year}{2019}), \eprint{1708.01527}.

\bibitem[{\citenamefont{Khalek et~al.}(2021)\citenamefont{Khalek, Ethier,
  Nocera, and Rojo}}]{Khalek:2021ulf}
\bibinfo{author}{\bibfnamefont{R.~A.} \bibnamefont{Khalek}},
  \bibinfo{author}{\bibfnamefont{J.~J.} \bibnamefont{Ethier}},
  \bibinfo{author}{\bibfnamefont{E.~R.} \bibnamefont{Nocera}},
  \bibnamefont{and} \bibinfo{author}{\bibfnamefont{J.}~\bibnamefont{Rojo}},
  \bibinfo{journal}{Phys. Rev. D} \textbf{\bibinfo{volume}{103}},
  \bibinfo{pages}{096005} (\bibinfo{year}{2021}), \eprint{2102.00018}.

\bibitem[{\citenamefont{Cruz-Martinez et~al.}(2024)\citenamefont{Cruz-Martinez,
  Fieg, Giani, Krack, M\"akel\"a, Rabemananjara, and
  Rojo}}]{Cruz-Martinez:2023sdv}
\bibinfo{author}{\bibfnamefont{J.~M.} \bibnamefont{Cruz-Martinez}},
  \bibinfo{author}{\bibfnamefont{M.}~\bibnamefont{Fieg}},
  \bibinfo{author}{\bibfnamefont{T.}~\bibnamefont{Giani}},
  \bibinfo{author}{\bibfnamefont{P.}~\bibnamefont{Krack}},
  \bibinfo{author}{\bibfnamefont{T.}~\bibnamefont{M\"akel\"a}},
  \bibinfo{author}{\bibfnamefont{T.~R.} \bibnamefont{Rabemananjara}},
  \bibnamefont{and} \bibinfo{author}{\bibfnamefont{J.}~\bibnamefont{Rojo}},
  \bibinfo{journal}{Eur. Phys. J. C} \textbf{\bibinfo{volume}{84}},
  \bibinfo{pages}{369} (\bibinfo{year}{2024}), \eprint{2309.09581}.

\bibitem[{\citenamefont{Boughezal et~al.}(2020)\citenamefont{Boughezal,
  Petriello, and Wiegand}}]{Boughezal:2020uwq}
\bibinfo{author}{\bibfnamefont{R.}~\bibnamefont{Boughezal}},
  \bibinfo{author}{\bibfnamefont{F.}~\bibnamefont{Petriello}},
  \bibnamefont{and} \bibinfo{author}{\bibfnamefont{D.}~\bibnamefont{Wiegand}},
  \bibinfo{journal}{Phys. Rev. D} \textbf{\bibinfo{volume}{101}},
  \bibinfo{pages}{116002} (\bibinfo{year}{2020}), \eprint{2004.00748}.

\bibitem[{\citenamefont{Bissolotti et~al.}(2023)\citenamefont{Bissolotti,
  Boughezal, and Simsek}}]{Bissolotti:2023pjh}
\bibinfo{author}{\bibfnamefont{C.}~\bibnamefont{Bissolotti}},
  \bibinfo{author}{\bibfnamefont{R.}~\bibnamefont{Boughezal}},
  \bibnamefont{and} \bibinfo{author}{\bibfnamefont{K.}~\bibnamefont{Simsek}}
  (\bibinfo{year}{2023}), \eprint{2307.09459}.

\bibitem[{\citenamefont{Aleksa et~al.}(2022)}]{Mangano:2022ukr}
\bibinfo{author}{\bibfnamefont{M.}~\bibnamefont{Aleksa}} \bibnamefont{et~al.},
  \textbf{\bibinfo{volume}{2/2022}} (\bibinfo{year}{2022}).

\bibitem[{\citenamefont{Ball et~al.}(2022{\natexlab{b}})\citenamefont{Ball,
  Candido, Forte, Hekhorn, Nocera, Rojo, and Schwan}}]{Ball:2022qtp}
\bibinfo{author}{\bibfnamefont{R.~D.} \bibnamefont{Ball}},
  \bibinfo{author}{\bibfnamefont{A.}~\bibnamefont{Candido}},
  \bibinfo{author}{\bibfnamefont{S.}~\bibnamefont{Forte}},
  \bibinfo{author}{\bibfnamefont{F.}~\bibnamefont{Hekhorn}},
  \bibinfo{author}{\bibfnamefont{E.~R.} \bibnamefont{Nocera}},
  \bibinfo{author}{\bibfnamefont{J.}~\bibnamefont{Rojo}}, \bibnamefont{and}
  \bibinfo{author}{\bibfnamefont{C.}~\bibnamefont{Schwan}},
  \bibinfo{journal}{Eur. Phys. J. C} \textbf{\bibinfo{volume}{82}},
  \bibinfo{pages}{1160} (\bibinfo{year}{2022}{\natexlab{b}}),
  \eprint{2209.08115}.

\bibitem[{\citenamefont{Anataichuk et~al.}(2023)}]{Anataichuk:2023elk}
\bibinfo{author}{\bibfnamefont{A.}~\bibnamefont{Anataichuk}}
  \bibnamefont{et~al.} (\bibinfo{year}{2023}), \eprint{2310.19638}.

\bibitem[{\citenamefont{Fiaschi et~al.}(2022)\citenamefont{Fiaschi, Giuli,
  Hautmann, and Moretti}}]{Fiaschi:2021sin}
\bibinfo{author}{\bibfnamefont{J.}~\bibnamefont{Fiaschi}},
  \bibinfo{author}{\bibfnamefont{F.}~\bibnamefont{Giuli}},
  \bibinfo{author}{\bibfnamefont{F.}~\bibnamefont{Hautmann}}, \bibnamefont{and}
  \bibinfo{author}{\bibfnamefont{S.}~\bibnamefont{Moretti}},
  \bibinfo{journal}{JHEP} \textbf{\bibinfo{volume}{02}}, \bibinfo{pages}{179}
  (\bibinfo{year}{2022}), \eprint{2111.09698}.

\bibitem[{\citenamefont{Aaron et~al.}(2010)}]{Aaron:2009aa}
\bibinfo{author}{\bibfnamefont{F.}~\bibnamefont{Aaron}} \bibnamefont{et~al.}
  (\bibinfo{collaboration}{H1 and ZEUS}), \bibinfo{journal}{JHEP}
  \textbf{\bibinfo{volume}{1001}}, \bibinfo{pages}{109} (\bibinfo{year}{2010}),
  \eprint{0911.0884}.

\bibitem[{\citenamefont{Abramowicz et~al.}(2013)}]{Abramowicz:1900rp}
\bibinfo{author}{\bibfnamefont{H.}~\bibnamefont{Abramowicz}}
  \bibnamefont{et~al.} (\bibinfo{collaboration}{H1 , ZEUS}),
  \bibinfo{journal}{Eur.Phys.J.} \textbf{\bibinfo{volume}{C73}},
  \bibinfo{pages}{2311} (\bibinfo{year}{2013}), \eprint{1211.1182}.

\bibitem[{\citenamefont{Abramowicz et~al.}(2015)}]{Abramowicz:2015mha}
\bibinfo{author}{\bibfnamefont{H.}~\bibnamefont{Abramowicz}}
  \bibnamefont{et~al.} (\bibinfo{collaboration}{ZEUS, H1}),
  \bibinfo{journal}{Eur. Phys. J.} \textbf{\bibinfo{volume}{C75}},
  \bibinfo{pages}{580} (\bibinfo{year}{2015}), \eprint{1506.06042}.

\bibitem[{\citenamefont{Arneodo et~al.}(1997)}]{Arneodo:1996qe}
\bibinfo{author}{\bibfnamefont{M.}~\bibnamefont{Arneodo}} \bibnamefont{et~al.}
  (\bibinfo{collaboration}{New Muon}), \bibinfo{journal}{Nucl. Phys.}
  \textbf{\bibinfo{volume}{B483}}, \bibinfo{pages}{3} (\bibinfo{year}{1997}),
  \eprint{hep-ph/9610231}.

\bibitem[{\citenamefont{Whitlow et~al.}(1992)\citenamefont{Whitlow, Riordan,
  Dasu, Rock, and Bodek}}]{Whitlow:1991uw}
\bibinfo{author}{\bibfnamefont{L.~W.} \bibnamefont{Whitlow}},
  \bibinfo{author}{\bibfnamefont{E.~M.} \bibnamefont{Riordan}},
  \bibinfo{author}{\bibfnamefont{S.}~\bibnamefont{Dasu}},
  \bibinfo{author}{\bibfnamefont{S.}~\bibnamefont{Rock}}, \bibnamefont{and}
  \bibinfo{author}{\bibfnamefont{A.}~\bibnamefont{Bodek}},
  \bibinfo{journal}{Phys. Lett.} \textbf{\bibinfo{volume}{B282}},
  \bibinfo{pages}{475} (\bibinfo{year}{1992}).

\bibitem[{\citenamefont{Benvenuti et~al.}(1989)}]{Benvenuti:1989rh}
\bibinfo{author}{\bibfnamefont{A.~C.} \bibnamefont{Benvenuti}}
  \bibnamefont{et~al.} (\bibinfo{collaboration}{BCDMS}),
  \bibinfo{journal}{Phys. Lett.} \textbf{\bibinfo{volume}{B223}},
  \bibinfo{pages}{485} (\bibinfo{year}{1989}).

\bibitem[{\citenamefont{Onengut et~al.}(2006)}]{Onengut:2005kv}
\bibinfo{author}{\bibfnamefont{G.}~\bibnamefont{Onengut}} \bibnamefont{et~al.}
  (\bibinfo{collaboration}{CHORUS}), \bibinfo{journal}{Phys. Lett.}
  \textbf{\bibinfo{volume}{B632}}, \bibinfo{pages}{65} (\bibinfo{year}{2006}).

\bibitem[{\citenamefont{Goncharov et~al.}(2001)}]{Goncharov:2001qe}
\bibinfo{author}{\bibfnamefont{M.}~\bibnamefont{Goncharov}}
  \bibnamefont{et~al.} (\bibinfo{collaboration}{NuTeV}),
  \bibinfo{journal}{Phys. Rev.} \textbf{\bibinfo{volume}{D64}},
  \bibinfo{pages}{112006} (\bibinfo{year}{2001}), \eprint{hep-ex/0102049}.

\bibitem[{\citenamefont{Abazov et~al.}(2008{\natexlab{a}})}]{Abazov:2008qv}
\bibinfo{author}{\bibfnamefont{V.~M.} \bibnamefont{Abazov}}
  \bibnamefont{et~al.} (\bibinfo{collaboration}{D0}), \bibinfo{journal}{Phys.
  Rev. Lett.} \textbf{\bibinfo{volume}{101}}, \bibinfo{pages}{211801}
  (\bibinfo{year}{2008}{\natexlab{a}}), \eprint{0807.3367}.

\bibitem[{\citenamefont{Aaltonen et~al.}(2011)}]{Aaltonen:2011gs}
\bibinfo{author}{\bibfnamefont{T.}~\bibnamefont{Aaltonen}} \bibnamefont{et~al.}
  (\bibinfo{collaboration}{CDF and D0}) (\bibinfo{year}{2011}),
  \eprint{1103.3233}.

\bibitem[{\citenamefont{Abazov et~al.}(2008{\natexlab{b}})}]{D0:2008hua}
\bibinfo{author}{\bibfnamefont{V.~M.} \bibnamefont{Abazov}}
  \bibnamefont{et~al.} (\bibinfo{collaboration}{D0}), \bibinfo{journal}{Phys.
  Rev. Lett.} \textbf{\bibinfo{volume}{101}}, \bibinfo{pages}{062001}
  (\bibinfo{year}{2008}{\natexlab{b}}), \eprint{0802.2400}.

\bibitem[{\citenamefont{Abazov et~al.}(2008{\natexlab{c}})}]{Abazov:2008qz}
\bibinfo{author}{\bibfnamefont{V.~M.} \bibnamefont{Abazov}}
  \bibnamefont{et~al.} (\bibinfo{collaboration}{D0}), \bibinfo{journal}{Phys.
  Lett.} \textbf{\bibinfo{volume}{B666}}, \bibinfo{pages}{23}
  (\bibinfo{year}{2008}{\natexlab{c}}), \eprint{0803.2259}.

\bibitem[{\citenamefont{Abazov et~al.}(2013)}]{Abazov:2013rja}
\bibinfo{author}{\bibfnamefont{V.~M.} \bibnamefont{Abazov}}
  \bibnamefont{et~al.} (\bibinfo{collaboration}{D0}),
  \bibinfo{journal}{Phys.Rev.} \textbf{\bibinfo{volume}{D88}},
  \bibinfo{pages}{091102} (\bibinfo{year}{2013}), \eprint{1309.2591}.

\bibitem[{\citenamefont{Abazov et~al.}(2015)}]{D0:2014kma}
\bibinfo{author}{\bibfnamefont{V.~M.} \bibnamefont{Abazov}}
  \bibnamefont{et~al.} (\bibinfo{collaboration}{D0}), \bibinfo{journal}{Phys.
  Rev.} \textbf{\bibinfo{volume}{D91}}, \bibinfo{pages}{032007}
  (\bibinfo{year}{2015}), \bibinfo{note}{[Erratum: Phys.
  Rev.D91,no.7,079901(2015)]}, \eprint{1412.2862}.

\bibitem[{\citenamefont{Aad et~al.}(2012{\natexlab{a}})}]{Aad:2011dm}
\bibinfo{author}{\bibfnamefont{G.}~\bibnamefont{Aad}} \bibnamefont{et~al.}
  (\bibinfo{collaboration}{ATLAS}), \bibinfo{journal}{Phys.Rev.}
  \textbf{\bibinfo{volume}{D85}}, \bibinfo{pages}{072004}
  (\bibinfo{year}{2012}{\natexlab{a}}), \eprint{1109.5141}.

\bibitem[{\citenamefont{Aaboud et~al.}(2017{\natexlab{a}})}]{Aaboud:2016btc}
\bibinfo{author}{\bibfnamefont{M.}~\bibnamefont{Aaboud}} \bibnamefont{et~al.}
  (\bibinfo{collaboration}{ATLAS}), \bibinfo{journal}{Eur. Phys. J.}
  \textbf{\bibinfo{volume}{C77}}, \bibinfo{pages}{367}
  (\bibinfo{year}{2017}{\natexlab{a}}), \eprint{1612.03016}.

\bibitem[{\citenamefont{Aad et~al.}(2014)}]{Aad:2014qja}
\bibinfo{author}{\bibfnamefont{G.}~\bibnamefont{Aad}} \bibnamefont{et~al.}
  (\bibinfo{collaboration}{ATLAS}), \bibinfo{journal}{JHEP}
  \textbf{\bibinfo{volume}{06}}, \bibinfo{pages}{112} (\bibinfo{year}{2014}),
  \eprint{1404.1212}.

\bibitem[{\citenamefont{Aad et~al.}(2013{\natexlab{a}})}]{Aad:2013iua}
\bibinfo{author}{\bibfnamefont{G.}~\bibnamefont{Aad}} \bibnamefont{et~al.}
  (\bibinfo{collaboration}{ATLAS}), \bibinfo{journal}{Phys.Lett.}
  \textbf{\bibinfo{volume}{B725}}, \bibinfo{pages}{223}
  (\bibinfo{year}{2013}{\natexlab{a}}), \eprint{1305.4192}.

\bibitem[{\citenamefont{Chatrchyan et~al.}(2012)}]{Chatrchyan:2012xt}
\bibinfo{author}{\bibfnamefont{S.}~\bibnamefont{Chatrchyan}}
  \bibnamefont{et~al.} (\bibinfo{collaboration}{CMS}),
  \bibinfo{journal}{Phys.Rev.Lett.} \textbf{\bibinfo{volume}{109}},
  \bibinfo{pages}{111806} (\bibinfo{year}{2012}), \eprint{1206.2598}.

\bibitem[{\citenamefont{Chatrchyan et~al.}(2014)}]{Chatrchyan:2013mza}
\bibinfo{author}{\bibfnamefont{S.}~\bibnamefont{Chatrchyan}}
  \bibnamefont{et~al.} (\bibinfo{collaboration}{CMS}),
  \bibinfo{journal}{Phys.Rev.} \textbf{\bibinfo{volume}{D90}},
  \bibinfo{pages}{032004} (\bibinfo{year}{2014}), \eprint{1312.6283}.

\bibitem[{\citenamefont{Chatrchyan
  et~al.}(2013{\natexlab{a}})}]{Chatrchyan:2013tia}
\bibinfo{author}{\bibfnamefont{S.}~\bibnamefont{Chatrchyan}}
  \bibnamefont{et~al.} (\bibinfo{collaboration}{CMS}), \bibinfo{journal}{JHEP}
  \textbf{\bibinfo{volume}{1312}}, \bibinfo{pages}{030}
  (\bibinfo{year}{2013}{\natexlab{a}}), \eprint{1310.7291}.

\bibitem[{\citenamefont{Khachatryan
  et~al.}(2016{\natexlab{a}})}]{Khachatryan:2016pev}
\bibinfo{author}{\bibfnamefont{V.}~\bibnamefont{Khachatryan}}
  \bibnamefont{et~al.} (\bibinfo{collaboration}{CMS}), \bibinfo{journal}{Eur.
  Phys. J.} \textbf{\bibinfo{volume}{C76}}, \bibinfo{pages}{469}
  (\bibinfo{year}{2016}{\natexlab{a}}), \eprint{1603.01803}.

\bibitem[{\citenamefont{Aaij et~al.}(2013)}]{Aaij:2012mda}
\bibinfo{author}{\bibfnamefont{R.}~\bibnamefont{Aaij}} \bibnamefont{et~al.}
  (\bibinfo{collaboration}{LHCb}), \bibinfo{journal}{JHEP}
  \textbf{\bibinfo{volume}{1302}}, \bibinfo{pages}{106} (\bibinfo{year}{2013}),
  \eprint{1212.4620}.

\bibitem[{\citenamefont{Aaij et~al.}(2015{\natexlab{a}})}]{Aaij:2015gna}
\bibinfo{author}{\bibfnamefont{R.}~\bibnamefont{Aaij}} \bibnamefont{et~al.}
  (\bibinfo{collaboration}{LHCb}), \bibinfo{journal}{JHEP}
  \textbf{\bibinfo{volume}{08}}, \bibinfo{pages}{039}
  (\bibinfo{year}{2015}{\natexlab{a}}), \eprint{1505.07024}.

\bibitem[{\citenamefont{Aaij et~al.}(2015{\natexlab{b}})}]{LHCb:2015kwa}
\bibinfo{author}{\bibfnamefont{R.}~\bibnamefont{Aaij}} \bibnamefont{et~al.}
  (\bibinfo{collaboration}{LHCb}), \bibinfo{journal}{JHEP}
  \textbf{\bibinfo{volume}{05}}, \bibinfo{pages}{109}
  (\bibinfo{year}{2015}{\natexlab{b}}), \eprint{1503.00963}.

\bibitem[{\citenamefont{Aaij et~al.}(2016{\natexlab{a}})}]{LHCb:2015mad}
\bibinfo{author}{\bibfnamefont{R.}~\bibnamefont{Aaij}} \bibnamefont{et~al.}
  (\bibinfo{collaboration}{LHCb}), \bibinfo{journal}{JHEP}
  \textbf{\bibinfo{volume}{01}}, \bibinfo{pages}{155}
  (\bibinfo{year}{2016}{\natexlab{a}}), \eprint{1511.08039}.

\bibitem[{\citenamefont{Aad et~al.}(2016{\natexlab{a}})}]{ATLAS:2016gic}
\bibinfo{author}{\bibfnamefont{G.}~\bibnamefont{Aad}} \bibnamefont{et~al.}
  (\bibinfo{collaboration}{ATLAS}), \bibinfo{journal}{JHEP}
  \textbf{\bibinfo{volume}{08}}, \bibinfo{pages}{009}
  (\bibinfo{year}{2016}{\natexlab{a}}), \eprint{1606.01736}.

\bibitem[{\citenamefont{Aaboud et~al.}(2017{\natexlab{b}})}]{ATLAS:2017rue}
\bibinfo{author}{\bibfnamefont{M.}~\bibnamefont{Aaboud}} \bibnamefont{et~al.}
  (\bibinfo{collaboration}{ATLAS}), \bibinfo{journal}{JHEP}
  \textbf{\bibinfo{volume}{12}}, \bibinfo{pages}{059}
  (\bibinfo{year}{2017}{\natexlab{b}}), \eprint{1710.05167}.

\bibitem[{\citenamefont{Aaij et~al.}(2016{\natexlab{b}})}]{Aaij:2016qqz}
\bibinfo{author}{\bibfnamefont{R.}~\bibnamefont{Aaij}} \bibnamefont{et~al.}
  (\bibinfo{collaboration}{LHCb}), \bibinfo{journal}{JHEP}
  \textbf{\bibinfo{volume}{10}}, \bibinfo{pages}{030}
  (\bibinfo{year}{2016}{\natexlab{b}}), \eprint{1608.01484}.

\bibitem[{\citenamefont{Aad et~al.}(2016{\natexlab{b}})}]{ATLAS:2016fij}
\bibinfo{author}{\bibfnamefont{G.}~\bibnamefont{Aad}} \bibnamefont{et~al.}
  (\bibinfo{collaboration}{ATLAS}), \bibinfo{journal}{Phys. Lett. B}
  \textbf{\bibinfo{volume}{759}}, \bibinfo{pages}{601}
  (\bibinfo{year}{2016}{\natexlab{b}}), \eprint{1603.09222}.

\bibitem[{\citenamefont{Aaij et~al.}(2016{\natexlab{c}})}]{Aaij:2016mgv}
\bibinfo{author}{\bibfnamefont{R.}~\bibnamefont{Aaij}} \bibnamefont{et~al.}
  (\bibinfo{collaboration}{LHCb}), \bibinfo{journal}{JHEP}
  \textbf{\bibinfo{volume}{09}}, \bibinfo{pages}{136}
  (\bibinfo{year}{2016}{\natexlab{c}}), \eprint{1607.06495}.

\bibitem[{\citenamefont{Aad et~al.}(2016{\natexlab{c}})}]{Aad:2015auj}
\bibinfo{author}{\bibfnamefont{G.}~\bibnamefont{Aad}} \bibnamefont{et~al.}
  (\bibinfo{collaboration}{ATLAS}), \bibinfo{journal}{Eur. Phys. J.}
  \textbf{\bibinfo{volume}{C76}}, \bibinfo{pages}{291}
  (\bibinfo{year}{2016}{\natexlab{c}}), \eprint{1512.02192}.

\bibitem[{\citenamefont{Khachatryan
  et~al.}(2015{\natexlab{a}})}]{Khachatryan:2015oaa}
\bibinfo{author}{\bibfnamefont{V.}~\bibnamefont{Khachatryan}}
  \bibnamefont{et~al.} (\bibinfo{collaboration}{CMS}), \bibinfo{journal}{Phys.
  Lett.} \textbf{\bibinfo{volume}{B749}}, \bibinfo{pages}{187}
  (\bibinfo{year}{2015}{\natexlab{a}}), \eprint{1504.03511}.

\bibitem[{\citenamefont{Aaboud et~al.}(2018)}]{ATLAS:2017irc}
\bibinfo{author}{\bibfnamefont{M.}~\bibnamefont{Aaboud}} \bibnamefont{et~al.}
  (\bibinfo{collaboration}{ATLAS}), \bibinfo{journal}{JHEP}
  \textbf{\bibinfo{volume}{05}}, \bibinfo{pages}{077} (\bibinfo{year}{2018}),
  \bibinfo{note}{[Erratum: JHEP 10, 048 (2020)]}, \eprint{1711.03296}.

\bibitem[{\citenamefont{Sirunyan et~al.}(2017)}]{CMS:2017gbl}
\bibinfo{author}{\bibfnamefont{A.~M.} \bibnamefont{Sirunyan}}
  \bibnamefont{et~al.} (\bibinfo{collaboration}{CMS}), \bibinfo{journal}{Phys.
  Rev. D} \textbf{\bibinfo{volume}{96}}, \bibinfo{pages}{072005}
  (\bibinfo{year}{2017}), \eprint{1707.05979}.

\bibitem[{\citenamefont{Aad et~al.}(2012{\natexlab{b}})}]{Aad:2011fc}
\bibinfo{author}{\bibfnamefont{G.}~\bibnamefont{Aad}} \bibnamefont{et~al.}
  (\bibinfo{collaboration}{ATLAS}), \bibinfo{journal}{Phys. Rev.}
  \textbf{\bibinfo{volume}{D86}}, \bibinfo{pages}{014022}
  (\bibinfo{year}{2012}{\natexlab{b}}), \eprint{1112.6297}.

\bibitem[{\citenamefont{Aad et~al.}(2013{\natexlab{b}})}]{Aad:2013lpa}
\bibinfo{author}{\bibfnamefont{G.}~\bibnamefont{Aad}} \bibnamefont{et~al.}
  (\bibinfo{collaboration}{ATLAS}), \bibinfo{journal}{Eur.Phys.J.}
  \textbf{\bibinfo{volume}{C73}}, \bibinfo{pages}{2509}
  (\bibinfo{year}{2013}{\natexlab{b}}), \eprint{1304.4739}.

\bibitem[{\citenamefont{Aad et~al.}(2015)}]{Aad:2014vwa}
\bibinfo{author}{\bibfnamefont{G.}~\bibnamefont{Aad}} \bibnamefont{et~al.}
  (\bibinfo{collaboration}{ATLAS}), \bibinfo{journal}{JHEP}
  \textbf{\bibinfo{volume}{02}}, \bibinfo{pages}{153} (\bibinfo{year}{2015}),
  \bibinfo{note}{[Erratum: JHEP09,141(2015)]}, \eprint{1410.8857}.

\bibitem[{\citenamefont{Chatrchyan
  et~al.}(2013{\natexlab{b}})}]{Chatrchyan:2012bja}
\bibinfo{author}{\bibfnamefont{S.}~\bibnamefont{Chatrchyan}}
  \bibnamefont{et~al.} (\bibinfo{collaboration}{CMS}),
  \bibinfo{journal}{Phys.Rev.} \textbf{\bibinfo{volume}{D87}},
  \bibinfo{pages}{112002} (\bibinfo{year}{2013}{\natexlab{b}}),
  \eprint{1212.6660}.

\bibitem[{\citenamefont{Khachatryan
  et~al.}(2016{\natexlab{b}})}]{Khachatryan:2015luy}
\bibinfo{author}{\bibfnamefont{V.}~\bibnamefont{Khachatryan}}
  \bibnamefont{et~al.} (\bibinfo{collaboration}{CMS}), \bibinfo{journal}{Eur.
  Phys. J.} \textbf{\bibinfo{volume}{C76}}, \bibinfo{pages}{265}
  (\bibinfo{year}{2016}{\natexlab{b}}), \eprint{1512.06212}.

\bibitem[{\citenamefont{Aaboud et~al.}(2017{\natexlab{c}})}]{ATLAS:2017kux}
\bibinfo{author}{\bibfnamefont{M.}~\bibnamefont{Aaboud}} \bibnamefont{et~al.}
  (\bibinfo{collaboration}{ATLAS}), \bibinfo{journal}{JHEP}
  \textbf{\bibinfo{volume}{09}}, \bibinfo{pages}{020}
  (\bibinfo{year}{2017}{\natexlab{c}}), \eprint{1706.03192}.

\bibitem[{\citenamefont{Khachatryan et~al.}(2017)}]{CMS:2016lna}
\bibinfo{author}{\bibfnamefont{V.}~\bibnamefont{Khachatryan}}
  \bibnamefont{et~al.} (\bibinfo{collaboration}{CMS}), \bibinfo{journal}{JHEP}
  \textbf{\bibinfo{volume}{03}}, \bibinfo{pages}{156} (\bibinfo{year}{2017}),
  \eprint{1609.05331}.

\bibitem[{\citenamefont{Aad et~al.}(2016{\natexlab{d}})}]{Aad:2016xcr}
\bibinfo{author}{\bibfnamefont{G.}~\bibnamefont{Aad}} \bibnamefont{et~al.}
  (\bibinfo{collaboration}{ATLAS}), \bibinfo{journal}{JHEP}
  \textbf{\bibinfo{volume}{08}}, \bibinfo{pages}{005}
  (\bibinfo{year}{2016}{\natexlab{d}}), \eprint{1605.03495}.

\bibitem[{\citenamefont{Aaboud et~al.}(2017{\natexlab{d}})}]{ATLAS:2017nah}
\bibinfo{author}{\bibfnamefont{M.}~\bibnamefont{Aaboud}} \bibnamefont{et~al.}
  (\bibinfo{collaboration}{ATLAS}), \bibinfo{journal}{Phys. Lett. B}
  \textbf{\bibinfo{volume}{770}}, \bibinfo{pages}{473}
  (\bibinfo{year}{2017}{\natexlab{d}}), \eprint{1701.06882}.

\bibitem[{\citenamefont{Adams et~al.}(1996)}]{e665}
\bibinfo{author}{\bibfnamefont{M.~R.} \bibnamefont{Adams}} \bibnamefont{et~al.}
  (\bibinfo{collaboration}{E665}), \bibinfo{journal}{Phys. Rev.}
  \textbf{\bibinfo{volume}{D54}}, \bibinfo{pages}{3006} (\bibinfo{year}{1996}).

\bibitem[{\citenamefont{Towell et~al.}(2001{\natexlab{a}})}]{Towell:2001nh}
\bibinfo{author}{\bibfnamefont{R.~S.} \bibnamefont{Towell}}
  \bibnamefont{et~al.} (\bibinfo{collaboration}{FNAL E866/NuSea}),
  \bibinfo{journal}{Phys. Rev.} \textbf{\bibinfo{volume}{D64}},
  \bibinfo{pages}{052002} (\bibinfo{year}{2001}{\natexlab{a}}),
  \eprint{hep-ex/0103030}.

\bibitem[{\citenamefont{Towell et~al.}(2001{\natexlab{b}})}]{NuSea:2001idv}
\bibinfo{author}{\bibfnamefont{R.~S.} \bibnamefont{Towell}}
  \bibnamefont{et~al.} (\bibinfo{collaboration}{NuSea}),
  \bibinfo{journal}{Phys. Rev. D} \textbf{\bibinfo{volume}{64}},
  \bibinfo{pages}{052002} (\bibinfo{year}{2001}{\natexlab{b}}),
  \eprint{hep-ex/0103030}.

\bibitem[{\citenamefont{Aad et~al.}(2016{\natexlab{e}})}]{Aad:2016zzw}
\bibinfo{author}{\bibfnamefont{G.}~\bibnamefont{Aad}} \bibnamefont{et~al.}
  (\bibinfo{collaboration}{ATLAS}), \bibinfo{journal}{JHEP}
  \textbf{\bibinfo{volume}{08}}, \bibinfo{pages}{009}
  (\bibinfo{year}{2016}{\natexlab{e}}), \eprint{1606.01736}.

\bibitem[{\citenamefont{Khachatryan et~al.}(2015{\natexlab{b}})}]{CMS:2014jea}
\bibinfo{author}{\bibfnamefont{V.}~\bibnamefont{Khachatryan}}
  \bibnamefont{et~al.} (\bibinfo{collaboration}{CMS}), \bibinfo{journal}{Eur.
  Phys. J.} \textbf{\bibinfo{volume}{C75}}, \bibinfo{pages}{147}
  (\bibinfo{year}{2015}{\natexlab{b}}), \eprint{1412.1115}.

\bibitem[{\citenamefont{Sirunyan et~al.}(2019)}]{CMS:2018mdl}
\bibinfo{author}{\bibfnamefont{A.~M.} \bibnamefont{Sirunyan}}
  \bibnamefont{et~al.} (\bibinfo{collaboration}{CMS}), \bibinfo{journal}{JHEP}
  \textbf{\bibinfo{volume}{12}}, \bibinfo{pages}{059} (\bibinfo{year}{2019}),
  \eprint{1812.10529}.

\bibitem[{\citenamefont{Abdul~Khalek et~al.}(2018)\citenamefont{Abdul~Khalek,
  Bailey, Gao, Harland-Lang, and Rojo}}]{Khalek:2018mdn}
\bibinfo{author}{\bibfnamefont{R.}~\bibnamefont{Abdul~Khalek}},
  \bibinfo{author}{\bibfnamefont{S.}~\bibnamefont{Bailey}},
  \bibinfo{author}{\bibfnamefont{J.}~\bibnamefont{Gao}},
  \bibinfo{author}{\bibfnamefont{L.}~\bibnamefont{Harland-Lang}},
  \bibnamefont{and} \bibinfo{author}{\bibfnamefont{J.}~\bibnamefont{Rojo}},
  \bibinfo{journal}{Eur. Phys. J.} \textbf{\bibinfo{volume}{C78}},
  \bibinfo{pages}{962} (\bibinfo{year}{2018}), \eprint{1810.03639}.

\bibitem[{\citenamefont{Iranipour and Ubiali}(2022)}]{Iranipour:2022iak}
\bibinfo{author}{\bibfnamefont{S.}~\bibnamefont{Iranipour}} \bibnamefont{and}
  \bibinfo{author}{\bibfnamefont{M.}~\bibnamefont{Ubiali}},
  \bibinfo{journal}{JHEP} \textbf{\bibinfo{volume}{05}}, \bibinfo{pages}{032}
  (\bibinfo{year}{2022}), \eprint{2201.07240}.

\bibitem[{\citenamefont{Ball et~al.}(2015)}]{NNPDF:2014otw}
\bibinfo{author}{\bibfnamefont{R.~D.} \bibnamefont{Ball}} \bibnamefont{et~al.}
  (\bibinfo{collaboration}{NNPDF}), \bibinfo{journal}{JHEP}
  \textbf{\bibinfo{volume}{04}}, \bibinfo{pages}{040} (\bibinfo{year}{2015}),
  \eprint{1410.8849}.

\bibitem[{\citenamefont{Del~Debbio et~al.}(2022)\citenamefont{Del~Debbio,
  Giani, and Wilson}}]{DelDebbio:2021whr}
\bibinfo{author}{\bibfnamefont{L.}~\bibnamefont{Del~Debbio}},
  \bibinfo{author}{\bibfnamefont{T.}~\bibnamefont{Giani}}, \bibnamefont{and}
  \bibinfo{author}{\bibfnamefont{M.}~\bibnamefont{Wilson}},
  \bibinfo{journal}{Eur. Phys. J. C} \textbf{\bibinfo{volume}{82}},
  \bibinfo{pages}{330} (\bibinfo{year}{2022}), \eprint{2111.05787}.

\bibitem[{\citenamefont{Alwall et~al.}(2014)\citenamefont{Alwall, Frederix,
  Frixione, Hirschi, Maltoni, Mattelaer, Shao, Stelzer, Torrielli, and
  Zaro}}]{Alwall:2014hca}
\bibinfo{author}{\bibfnamefont{J.}~\bibnamefont{Alwall}},
  \bibinfo{author}{\bibfnamefont{R.}~\bibnamefont{Frederix}},
  \bibinfo{author}{\bibfnamefont{S.}~\bibnamefont{Frixione}},
  \bibinfo{author}{\bibfnamefont{V.}~\bibnamefont{Hirschi}},
  \bibinfo{author}{\bibfnamefont{F.}~\bibnamefont{Maltoni}},
  \bibinfo{author}{\bibfnamefont{O.}~\bibnamefont{Mattelaer}},
  \bibinfo{author}{\bibfnamefont{H.~S.} \bibnamefont{Shao}},
  \bibinfo{author}{\bibfnamefont{T.}~\bibnamefont{Stelzer}},
  \bibinfo{author}{\bibfnamefont{P.}~\bibnamefont{Torrielli}},
  \bibnamefont{and} \bibinfo{author}{\bibfnamefont{M.}~\bibnamefont{Zaro}},
  \bibinfo{journal}{JHEP} \textbf{\bibinfo{volume}{07}}, \bibinfo{pages}{079}
  (\bibinfo{year}{2014}), \eprint{1405.0301}.

\end{thebibliography}

\onecolumngrid
\appendix

\section{Estimators and list of data and pseudo-data}
\label{app:datasets}
In this Appendix, we provide additional information about the methodology and the definition of the statistical 
estimators used in our analysis. We also provide a list of experimental data and pseudo-data that we 
include, as well as details on the BSM model that we consider throughout the paper. 
Additionally we show how our findings vary as a weaker BSM signal is injected in the data. We investigate 
what is the maximal BSM signal that may be absorbed in PDF fit even in the presence of EIC and FPF data and 
what the effect would be on indirect BSM searches. 

\paragraph{Closure test.}
In order to systematically study possible BSM-induced bias in PDF fits, we work in a setting 
in which we assume we know the underlying law of nature. In our case the law of nature 
consists of the {\it true} PDFs, which are low-energy quantities that have nothing to do with 
new physics, and the {\it true} UV complete Lagrangian, which at low energy is well approximated 
by the SM Lagrangian but which comprises heavy new particles. 
We use these assumptions to generate the pseudo-data in our analysis.
We inject the effect of the new particles that we introduce in the Lagrangian 
in the articifially generated data, and their effect 
will be visible in some high-energy distributions depending on the underlying model. 
The methodology we use throughout this study is based on the NNPDF \textit{closure-test} framework,
first introduced in Ref.~\cite{NNPDF:2014otw}, and explained in more detail in 
Ref.~\cite{DelDebbio:2021whr}. Various statistical estimators are applied to check 
the quality of the fit (in broad terms, assessing its difference 
from the {\it true} PDFs), hence verifying the accuracy of the fitting methodology.

\paragraph{Detection of inconsistencies in global PDF fits.}
Once we produce a closure test fit, in which PDFs have been fitted assuming the SM but the 
artificial data have been produced according to a given BSM model we check whether it is possible 
to detect BSM-induced bias in the PDF fit, 
i.e. whether the datasets which \textit{are} affected by he given BSM appear inconsistent with the bulk of the 
data in the global analysis and 
are poorly described by the resulting PDFs.
To quantitatively address this point, we use the NNPDF dataset selection criteria, 
discussed in detail in~\cite{NNPDF:2021njg}.  
We consider both the $\chi^2$-statistic of the resulting fit to each dataset 
entering the fit, and also consider the number of standard deviations 
\begin{equation}
n_{\sigma} = (\chi^2 - N_{\rm dat}) / \sqrt{2N_{\rm dat}}
\end{equation} 
of the $\chi^2$-statistics from the expected $\chi^2$ for each dataset. 
If $\chi^2/N_{\rm dat} > 1.5$ and $n_{\sigma} > 2$ for a particular dataset, the dataset would be 
flagged by the NNPDF selection criteria, indicating an inconsistency with the other data entering 
the fit.

\paragraph{Weighted fits.}
There are two possible outcomes from performing such a dataset selection analysis on a fit. 
In the first instance, the datasets affected by NP are flagged by the dataset selection criterion. 
If a dataset is flagged according to this condition, then a weighted fit is performed, 
{\it i.e.} a fit in which a dataset $(j)$ is given a
larger weight inversely proportional to the number of datapoints,
namely
\begin{equation}
\chi^2=\frac{1}{N_{\rm dat}}\,\sum_{i=1}^{N_{\rm dat}} \,N_{\rm
  dat}^{(i)}\,\chi^2_i , \qquad \chi^2_w=\frac{1}{N_{\rm dat}-N_{\rm
    dat}^{(j)}}\,\sum_{i=1,i\neq j}^{n_{\rm exp}} \,N_{\rm
  dat}^{(i)}\,\chi^2_i \, + \,w^{(j)}\chi^2_j,
  \end{equation}
  with
  \begin{equation}
    \label{eq:wgts}
    w^{(j)} = N_{\rm dat}/N_{\rm dat}^{(j)}.
    \end{equation}
If the data-theory agreement improves by setting it below thresholds and the
data-theory agreement of the other datasets does not deteriorate in
any statistically relevant way, then the dataset is kept, else the dataset is discarded, 
on the basis of the inconsistency with the remaining datasets.

\paragraph{Statistical distributions.} 
Moreover, to check if the flag raised in the post-fit analysis is statistically 
significant, we repeat the $\chi^{2(k)}$ and $n_\sigma^{(k)}$ computations 
across 1000 random instances $(k)$ of the data generation in the closure-test fits 
(corresponding to independent runs of the Universe). 
We check it both globally and for each individual datasets, with a special 
emphasis on the datasets that are mostly affected by the BSM corrections 
associated to the given model. If the $\chi^2$ and $n_{\sigma}$ distributions are peaked above the critical 
thresholds of 1.5 and 2 respectively, then the weighted fit procedure is 
activated and the inconsistency is assessed according to the methodology outlined above. 

\paragraph{BSM model.}
The model that we consider thoughout this paper consists of the $SU(2)_L$ triplet 
field ${W'}^{a, \mu}$, where $a \in \{1,2,3\}$ denotes an $SU(2)_L$ index, with mass 
$M_{W'}$ and coupling coefficient by $g_{W'}$. The model is explained in details in 
Refs.~\cite{Farina:2016rws,Greljo:2021kvv} and is described by the Lagrangian:
\begin{equation} \label{eq:Wprime}
	\begin{split}
		\mathcal{L}^{W'}_{\text{UV}} &= \mathcal{L}_{\text{SM}} - \frac{1}{4} {W'}^{a}_{\mu \nu} {W'}^{a, \mu \nu} + \frac{1}{2} M_{W'}^{2} {W'}_{\mu}^{a} {W'}^{a, \mu}\\
		&\qquad - g_{W'} {W'}^{a,\mu} \sum_{\substack{f_L}} \bar{f}_L T^{a} \gamma^{\mu} f_L
		 - g_{W'} ({W'}^{a, \mu} \varphi^{\dagger} T^{a} i D_{\mu} \varphi + \textrm{h.c.}) \, ,
	\end{split}
\end{equation}
where the sum runs over the left-handed fermions: $f_L \in \{q, \ell\}$. The $SU(2)_L$ generators 
are given by $T^a = \frac{1}{2} \sigma ^a$ where $\sigma ^a$ are the Pauli matrices. The kinetic term 
is given by $W^{'a}_{\mu \nu} = \partial_{\mu} {W'}^{a}_{\nu} - \partial_{\nu} {W'}^{a}_{\mu}- ig_{W'} [{W'}^{a}_{\mu}, {W'}^{a}_{\nu} ]$. The covariant derivative is given by
$D_{\mu} = \partial_{\mu} + \frac{1}{2} ig \sigma^{a} W_{\mu}^{a} + i g^{'} Y_{\varphi} B_{\mu}$.
The mixing with the SM gauge fields is neglected. The leading effect of this model on the dataset 
included in our analysis is to modify the Drell-Yan and deep inelastic scattering observables,
both charged current and neutral current. 
The tree-level matching of $\mathcal{L}^{W'}_{\text{UV}}$ to the dimension-6 operators in the SMEFT is described in details in 
Ref.~\cite{Greljo:2021kvv}. There it is shown that the impact at the LHC energy is dominated by the four-fermion
interactions, which sum to:
\begin{equation}
	\mathcal{L}^{W'}_{\text{SMEFT}} = \mathcal{L}_{\text{SM}} -\frac{g_{W'}^{2}}{2 M^2_{W'}} J^{a, \mu}_L J^a_{L, \mu}, \qquad J^{a, \mu}_L = \sum_{\substack{f_L}} \bar{f_L} T^a \gamma^{\mu} f_L \, ,
\end{equation}
which is typically described by the $\hat{W}$ parameter:
\begin{equation}
	\mathcal{L}^{W'}_{\text{SMEFT}} = \mathcal{L}_{\text{SM}} -\frac{g^{2} \hat{W}}{2 m^2_W} J^{a, \mu}_L J^a_{L, \mu}, \qquad \hat{W} = \frac{g_{W'}^{2}}{g^{2}} \frac{m_{W}^{2}}{M_{W'}^{2}} \, .
\end{equation}
Using Fermi's constant, one can write the relation between the UV parameters and $\hat{W}$ as
\begin{equation}
	\frac{g_{W'}^{2}}{M_{W'}^{2}} = 4 \sqrt{2} G_{F} \hat{W} \, .
\end{equation}
By fixing $g_{W'}=1$, each $M_{W'}$ can be associated to a value of $\hat{W}$.

\paragraph{Datasets included.}
We use all datasets included in the NNPDF4.0 global analysis~\cite{NNPDF:2021njg}.
The high-mass Drell-Yan measurements from ATLAS and CMS are impacted by the $W'$ model we consider, 
we provide details on the datapoints in Tab.~\ref{tab:HMDY_data}. Several fixed-target Drell-Yan measurements 
from Fermilab with proton (p) and deuteron (d) targets probe the same processes at much lower energies. The effects of new physics are much more suppressed
and they remain compatible with the SM predictions. More details is provided in Tab.~\ref{tab:FT_data}.
On top of the existing measurements, we include the high-mass Drell-Yan projections that were introduced in~\cite{Greljo:2021kvv}, 
inspired by the HL-LHC projections studied in~\cite{Khalek:2018mdn}.
The invariant mass distribution projections are generated at 14 TeV assuming an integrated 
luminosity of 6 ab$^{-1}$ (3 ab$^{-1}$ collected by ATLAS and 3 ab$^{-1}$ by CMS). Both in 
the case of NC and CC Drell-Yan cross sections, the MC data were generated using the
MadGraph5 aMCatNLO NLO Monte Carlo event generator~\cite{Alwall:2014hca} with additional $K$-factors to
include the NNLO QCD and NLO EW corrections. The MC data consist of four datasets
(associated with NC/CC distributions with muons/electrons in the final state), each comprising 
16 bins in the $M_{ll}$ invariant mass distribution or transverse mass $M_T$ distributions
with both $m_{ll}$ and $M_T$ greater than 500 GeV , with the highest energy bins reaching $m_{ll}$ = 4
TeV ($M_T$ = 3.5 TeV) for NC (CC) data. The rationale behind the choice of number of bins
and the width of each bin was outlined in Ref.~\cite{Greljo:2021kvv}, and stemmed from the requirement
that the expected number of events per bin was big enough to ensure the applicability of
Gaussian statistics. The choice of binning for the $M_{ll} (M_T)$ distribution at the HL-LHC is
displayed in Fig.~5.1 of Ref.~\cite{Greljo:2021kvv}. The details of the datasets included in the fit are 
presented in Tab.~\ref{tab:HLLHC_data}.
\begin{table}[ht]
  \centering
  \setlength{\tabcolsep}{2pt} 
  \renewcommand{\arraystretch}{1.5} 
  \begin{tabularx}{\textwidth}{X X X}
      \begin{minipage}{\linewidth}
          \centering
          \small
          \begin{tabular}{|c|c|c|}
\hline
\textbf{Observable} $\quad$ & $N_{\rm dat}$ & $\sqrt{s}$ [TeV] \\
\hline
\textbf{ATLAS} & & \\
$d\sigma^Z/dM_{\ell\ell}$ & 13 & 7\\
$d^2\sigma^Z/dM_{\ell\ell}d|y_{\ell\ell}|$ & 48 & 8 \\
 \hline
 \textbf{CMS} & & \\
 $d\sigma^Z/dy_{\mu^+ \mu^-}$ & 132 & 7 \\
 $d\sigma^Z/dy_{e^+ e^-}$ & 132 & 7 \\
 $d\sigma^Z/dM_{\ell\ell}$ & 43 & 13\\
 \hline
\end{tabular}

          \captionof{table}{\small Summary of the high-mass Drell-Yan data included in NNPDF4.0~\cite{NNPDF:2021njg}.}
          \label{tab:HMDY_data}
      \end{minipage} &
      \begin{minipage}{\linewidth}
          \centering
          \small
          \begin{tabular}{|c|c|c|}
\hline
\textbf{Observable} $\quad$ & $N_{\rm dat}$ & $\sqrt{s}$ [GeV] \\
\hline
\textbf{NuSea (E866)} & & \\
$d\sigma^{Z/\gamma^{*}}/dM_{ll}$ (d)& 15 & 800 \\
$d\sigma_{Z/\gamma^{*}}/dM_{ll}$ (p) & 184 & 800 \\
\hline
\textbf{E605} & & \\
$d\sigma^{Z/\gamma^{*}}/dM_{ll}$ & 119 & 38.8 \\
\hline
\textbf{SeaQuest (E905)} & & \\
$d\sigma^{Z/\gamma^{*}}/dM_{ll}$ (d) & 6 & 120 \\
 \hline
\end{tabular}

          \captionof{table}{\small Summary of the fixed-target Drell-Yan data included in NNPDF4.0~\cite{NNPDF:2021njg}.}
          \label{tab:FT_data}
      \end{minipage} &
      \begin{minipage}{\linewidth}
          \centering
          \small
          \begin{tabular}{|c|c|c|}
\hline
\textbf{Observable} $\quad$ & $N_{\rm dat}$ & $\sqrt{s}$ [TeV]\\
\hline
\textbf{Charged Current} & & \\
$d\sigma^{W^\pm}/dM_{T,e\nu}$ & 16 & 14 \\
$d\sigma^{W^\pm}/dM_{T,\mu\nu}$ & 16 & 14 \\
\hline
\textbf{Neutral Current} & & \\
$d\sigma^{Z}/dM_{e^+e^-}$ & 12 & 14 \\
$d\sigma^{Z}/dM_{\mu^+\mu^-}$ & 12 & 14 \\
\hline
\end{tabular}

          \captionof{table}{\small Summary of the HL-LHC projected data included in this work, taken from Ref.~\cite{Greljo:2021kvv}.}
          \label{tab:HLLHC_data}
      \end{minipage}
  \end{tabularx}
\end{table}

As far as the EIC projections are concerned, we consider those that were built in  
Ref.~\cite{Khalek:2021ulf}, in particular the measurements of charged-current (CC) and 
neutral-current (NC) DIS processes, using both electrons and positrons as projectiles and 
both protons and neutrons as targets. Details about the projections used are presented in 
Tab.~\ref{tab:EIC_data}. We tested both the optimistic and pessimistic scenarios which vary in terms of
how many datapoints can be collected. We found that the choice of scenario does not impact much our results.

For the FPF data, we consider the projected data built in Ref.~\cite{Cruz-Martinez:2023sdv}. 
The focus is on charged-current DIS processes, where the projectiles are neutrinos
and anti-neutrinos, coming from the LHC beam, hitting isoscalar targets and the processes associated 
with charm production. For each of the far-forward LHC neutrino experiments  
the pseudo-rapidity coverage, target material, the acceptance for the charged lepton and hadronic final state, and the expected
reconstruction performance are identified in~\cite{Cruz-Martinez:2023sdv} 
assuming that FASER$\nu$ and SND acquire data for Run III with 
${\cal L}$ = 150 fb$^{-1}$. SND projections are too statistically insignificant to be included in the fit, 
due to the little statistics collected so far. As far as the FPF facility is concerned, 
FASER$\nu$2, AdvSND, and FLArE take data for the complete 
HL-LHC period, with ${\cal L}$ = 3 ab$^{-1}$. The complete set of projected data that is included is 
summarised in Tab.~\ref{tab:FPF_data}.

\begin{table}[ht]
  \centering
  \setlength{\tabcolsep}{5pt} 
  \renewcommand{\arraystretch}{1.3} 
  \begin{tabularx}{\textwidth}{X X}
      \begin{minipage}{\linewidth}
          \centering
          \begin{tabular}{|c|c|c|c|}
\hline
\textbf{Observable} $\quad$ & $N_{dat}$ & $\sqrt{s}$ [GeV] & $\mathcal{L}$ [fb$^-1$] \\

\hline
\textbf{Charged Current} & & & \\
$\tilde{\sigma}(e^- + p \rightarrow \nu + X)$ & 89 (89) & 140.7 & 100 \\
$\tilde{\sigma}(e^+ +p \rightarrow \bar{\nu} + X)$ & 89 (89) & 140.7 & 10 \\

 \hline
 \textbf{Neutral Current (proton)} & & &\\
$\tilde{\sigma}(e^- + p \rightarrow e^- + X)$ & 181 (131) & 140.7 & 100 \\
& 181 (131) & 63.2 & 100 \\
& 126 (91) & 44.7 & 100 \\
& 87 (76) & 28.6 & 100 \\
$\tilde{\sigma}(e^+ + p \rightarrow e^+ + X)$ & 181 (131) & 140.7 & 10 \\
& 181 (131) & 63.2 & 10 \\
& 126 (91) & 44.7 & 10 \\
& 87 (76) & 28.6 & 10 \\

 \hline
 \textbf{Neutral Current (deuteron)} & & &\\
$\tilde{\sigma}(e^- + d \rightarrow e^- + X)$ & 116 (116) & 89.0 & 10 \\
& 107 (107) & 66.3 & 10 \\
& 76 (76) & 28.6 & 10 \\
$\tilde{\sigma}(e^+ + d \rightarrow e^+ + X)$& 116 (116) & 89.0 & 10 \\
& 107 (107) & 66.3 & 10 \\
& 76 (76) & 28.6 & 10 \\

 \hline

\end{tabular}

          \captionof{table}{Summary of the EIC projected data included in this work, taken from Ref.~\cite{Khalek:2021ulf}.}
          \label{tab:EIC_data}
      \end{minipage} &
      \begin{minipage}{\linewidth}
          \centering
          \renewcommand{\arraystretch}{1.5}
              \begin{tabular}{|c|c|}
        \hline
        \textbf{Observable} $\quad$ & $N_{dat}$\\
        
        \hline
        \textbf{FASER$\nu$} & \\
        $\tilde{\sigma}(\nu + N \rightarrow l^- + X)$ & 22  \\
        $\tilde{\sigma}(\bar{\nu} + N \rightarrow l^+ + X)$ & 16  \\

        \hline
        \textbf{FLArE} & \\
        $\tilde{\sigma}(\nu + N \rightarrow l^- + X)$ & 43  \\
        $\tilde{\sigma}(\bar{\nu} + N \rightarrow l^+ + X)$ & 39  \\
        $\tilde{\sigma}(\nu + N \rightarrow l^- + c/\bar{c} + X)$ & 31  \\
        $\tilde{\sigma}(\bar{\nu} + N \rightarrow l^+ + c/\bar{c} + X)$ & 19  \\
        
        \hline
        \textbf{FASER$\nu$2} & \\
        $\tilde{\sigma}(\nu + N \rightarrow l^- + X)$ & 44  \\
        $\tilde{\sigma}(\bar{\nu} + N \rightarrow l^+ + X)$ & 39  \\
        $\tilde{\sigma}(\nu + N \rightarrow l^- + c/\bar{c} + X)$ & 38  \\
        $\tilde{\sigma}(\bar{\nu} + N \rightarrow l^+ + c/\bar{c} + X)$ & 30  \\

        \hline
        \textbf{AdvSND} & \\
        $\tilde{\sigma}(\nu + N \rightarrow l^- + X)$ & 33  \\
        $\tilde{\sigma}(\bar{\nu} + N \rightarrow l^+ + X)$ & 29  \\
        $\tilde{\sigma}(\nu + N \rightarrow l^- + c/\bar{c} + X)$ & 17  \\

        \hline
    
    \end{tabular}
    
          \captionof{table}{Summary of the FPF projected data included in this work, taken from Ref.~\cite{Cruz-Martinez:2023sdv}.}
          \label{tab:FPF_data}
      \end{minipage}
  \end{tabularx}
\end{table}

\begin{figure}[htb]
   \begin{center}
       \setlength{\tabcolsep}{5pt} 
   \renewcommand{\arraystretch}{1.5} 
     \makebox{\includegraphics[width=0.45\columnwidth]{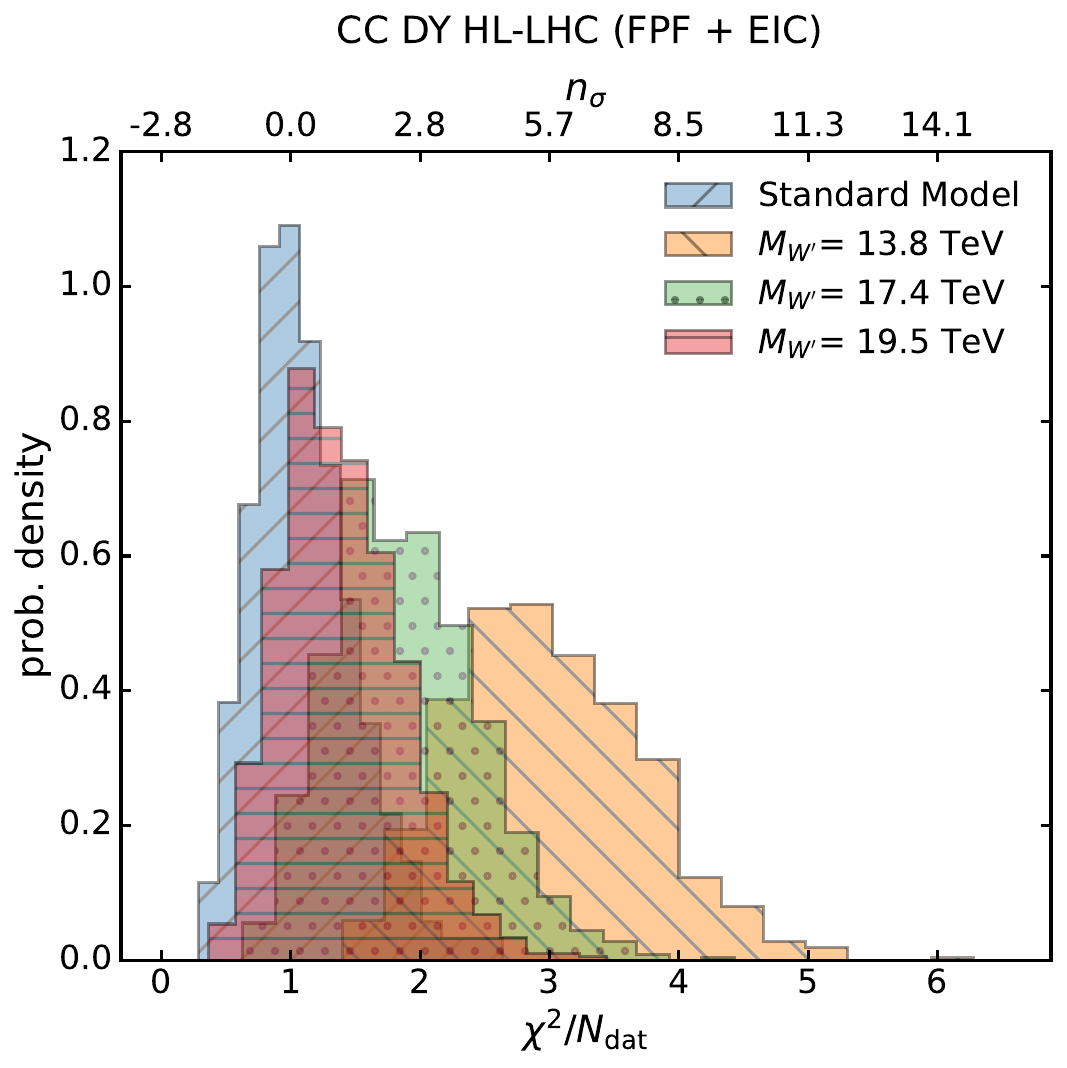}}
   \end{center}
   \vspace{-0.5cm}
   \caption{\small Comparison of the distribution of the $\chi^2$ per datapoint (bottom $x$-axis) and 
   $n_\sigma$ (top $x$-axis) for the high-mass HL-LHC DY CC projected data between
   the baseline fit (blue histogram) and fits with HL-LHC+EIC+FPF projections included by varying the 
   strenght of the BSM induced signal as $M_{W'}$ is increased from $M_{W'}=13.8$ TeV (red histogram) to 
   $M_{W'}=17.4$ TeV (green histogram) to $M_{W'}=19.5$ TeV (orange histogram).}
   \label{fig:chi2_EIC+FPF}
 \end{figure}

\paragraph{Maximal BSM threshold.}
To conclude, we consider how the maximal BSM threshold, {\it i.~e.} the maximal BSM signal that can be 
absorbed by PDFs in a global fit, shifts as we include FPF and EIC projected data. 
Clearly, if we inject in the data the effect of a new heavy flavour-universal $W'$, 
with mass larger than $13.8$ TeV, the depletion in the high-energy tails measured at the HL-LHC will 
be less significant and thus the effect can still be fitted away in the PDFs. In this appendix we 
investigate by how much the EIC and CERN neutrino data would shift the maximal BSM 
threshold and whether the impact of the BSM bias - pushed at a higher threshold - would visibly 
impact the PDFs and thus possibly hamper the interpretation of indirect new physics searches. 
To answer this question, we made a scan over $M_{W'}$, by injecting increasingly large values of 
the $W'$ mass, until the signal 
can be absorbed even in the presence of precise EIC and CERN neutrino data in a PDF global analysis. 
The summary results of the analysis are displayed in Fig.~\ref{fig:chi2_EIC+FPF}. There we see that, 
as we increase the value 
of $M_{W'}$ in our BSM model, thus decrease the size of the BSM corrections injected in the data, we 
have to get 
to $M_{W'}=19.5$ TeV (orange histogram) for the signal to be absorbed in the PDFs. 
\begin{figure*}[hbt]
  \begin{center}
    \makebox{\includegraphics[width=0.47\textwidth]{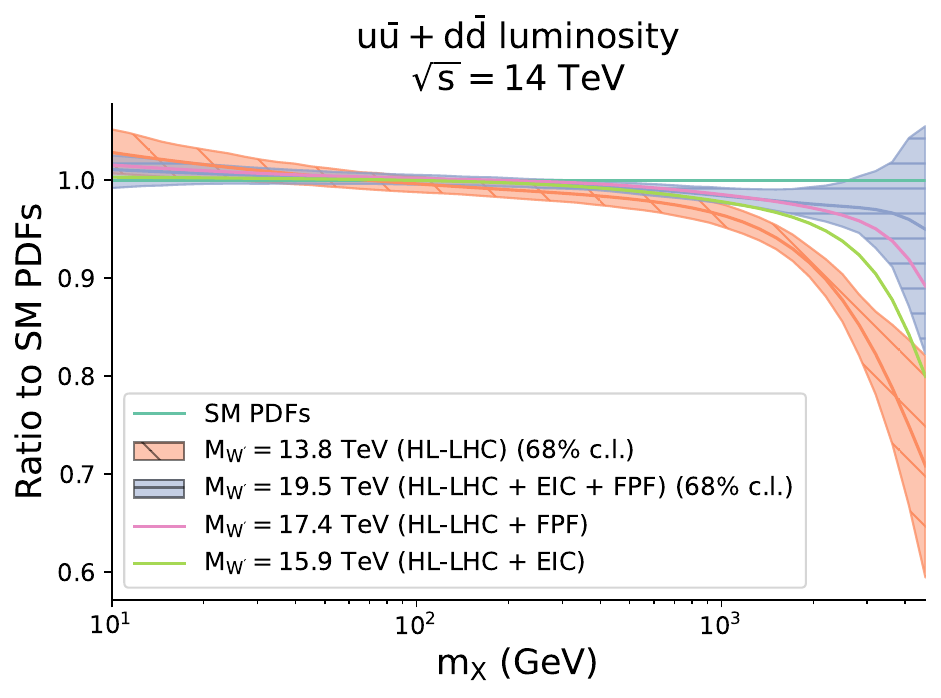}}
    \makebox{\includegraphics[width=0.45\textwidth]{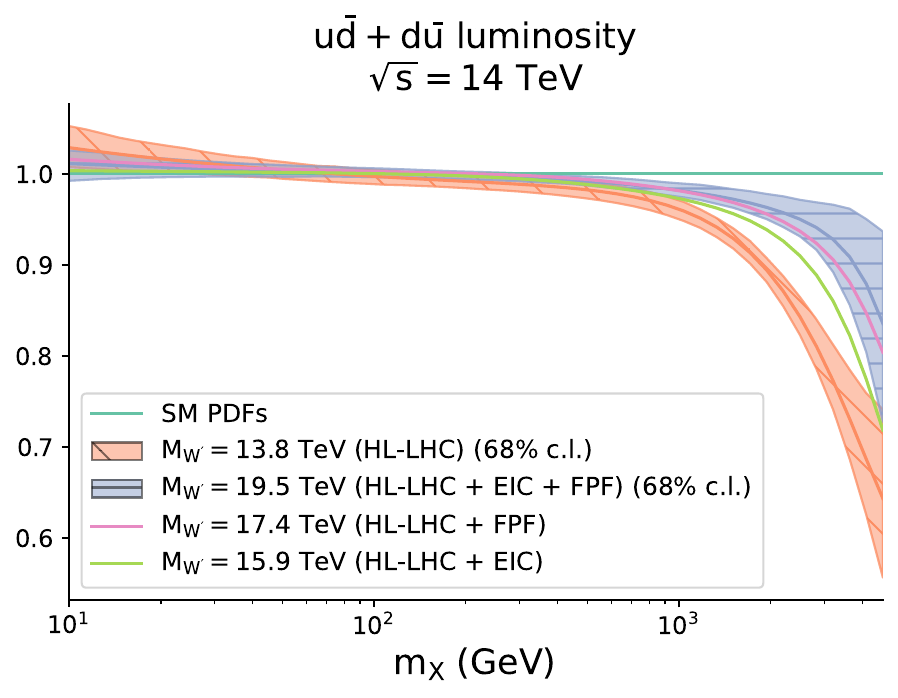}}
  \end{center}
  \vspace{-0.5cm}
  \caption{\small Impact of the BSM-induced bias on the NC $u\bar{u}+d\bar{d}$ luminosity, Eq.~\eqref{eq:zlumi}, (left) and 
  CC $u\bar{d}+d\bar{u}$ luminosity, Eq.~\eqref{eq:wlumi}, (right) for the different scenarios, all normalised to the 
  {\it true} SM PDF luminosities. The orange bands correspond to the central value and PDF uncertainties of a fit 
  including the global NNPDF4.0 datasets plus the HL-LHC projections considered here. The   
data have been generated according to a $W'$ model with $M_{W'}=13.8$ TeV and the fit is done assuming the SM. 
The blue bands correspond to the same fit in which also the EIC and the CERN forward neutrino projected data 
are included alongside the HL-LHC one, thus moving $M_{W'}$ to 19.5 TeV. 
The intermediate cases in which only EIC projections or only FPF projections are 
included are shown with only their central values in green and pink respectively, 
according to the maximal BSM signal that can be absorbed in each of these cases, namely $M_{W'}=15.9$ TeV and 
$M_{W'}=17.4$ TeV.}
  \label{fig:maxBSM}
\end{figure*}
Finally, we ask ourselves what is the effect of the new physics absorption now that the threshold has been pushed 
to a significantly larger value. Is this still going to spoil the interpretation of indirect BSM searches?
To answer this question we assess the impact of the injection of our new physics model in the data on the 
fitted PDFs, by looking at the integrated luminosities for the parton pair $i,j$, which is defined as:
\begin{equation}
\label{eq:integrated_luminosity}
  {\cal L}_{ij}(m_X,\sqrt{s}) = \frac{1}{s}\int\limits_{-y}^{y}\,d\tilde{y}\,
  \left[
  f_{i}\left(\frac{m_X}{\sqrt{s}}e^{\tilde{y}},m_X\right)\, 
  f_j\left(\frac{m_X}{\sqrt{s}}e^{-\tilde{y}},m_X\right)
  +
  (i \leftrightarrow j)
  \right]
  ,
  \end{equation}
 where $f_i \equiv f_i(x,Q)$ is the PDF corresponding to the parton flavour $i$ 
 , and the integration limits are defined by:
 \begin{equation}
 \label{eq:rapidity}
  y=\ln\left(\frac{\sqrt{s}}{m_X}\right).
  \end{equation}
In particular, we focus on the luminosities that are most constrained by the NC and CC Drell-Yan data respectively, namely   
  \begin{align}
    {\cal L}^{\rm NC}(m_X,\sqrt{s}) &= {\cal L}_{u\bar{u}}(m_X,\sqrt{s}) + {\cal L}_{d\bar{d}}(m_X,\sqrt{s}),\label{eq:zlumi}\\[1.5ex]
      {\cal L}^{\rm CC}(m_X,\sqrt{s}) &= {\cal L}_{u\bar{d}}(m_X,\sqrt{s}) + {\cal L}_{d\bar{u}}(m_X,\sqrt{s}).\label{eq:wlumi}
    \end{align}
In Fig.~\ref{fig:maxBSM} we show how the addition of the new projected data mitigates the BSM-induced PDF bias. 
We compare the NC and CC PDF luminosities defined in Eqs.~\eqref{eq:zlumi} and~\eqref{eq:wlumi} respectively 
as a function of $m_X$, which in the case 
of NC can be identified with the invariant mass of the produced leptons and in the case of CC can be identified with the 
transverse mass of the lepton-neutrino pair. We compare the luminosities obtained in a closure test in which the data 
have been generated by injecting a $W'$ model with $M_{W'}$ set at different values and the data fitted assuming the SM. 
Each of the curves is compared to the {\it true} SM luminosities. The orange band represents the luminosities fitted on 
the whole NNPDF4.0 dataset plus the HL-LHC projected data displayed in Table~\ref{tab:HLLHC_data} with a $M_{W'}=13.8$ TeV 
signal injected in the data and the fit performed assuming the SM. 
Here the result is the same as the one presented in Ref.~\cite{Hammou:2023heg}, namely 
the luminosities shift significantly as the PDFs have completely absorbed the BSM signal. On the other hand, if the 
EIC and the CERN forward neutrino projected data is included alongside the HL-LHC one, then the maximal BSM signal that can be absorbed 
corresponds to $M_{W'}=19.5$ TeV. The corresponding PDF luminosities are displayed in blue. 
We observe a systematic reduction of the discrepancy 
with the true SM luminosities. In the case of the $u\bar{u}+d\bar{d}$ luminosity, the fit including EIC and CERN forward neutrino 
projections almost completely suppresses the discrepancy while for the $u\bar{d}+d\bar{u}$ a visible  
shift remains but its effect is greatly reduced. The intermediate case in which only EIC projections or only FPF projections are 
included are shown with only their central values in green and pink respectively, 
according to the maximal BSM signal that can be absorbed in each of these cases.
To conclude, if the EIC and FPF projections are included in a global PDF analysis alongside the HL-LHC 
high-mass Drell-Yan distributions, the maximal BSM threshold, {\it i.~e.} the maximal BSM signal that can be 
absorbed by PDFs in a global fit, shifts to larger values. As a result, while the effect can still be fitted away in the PDFs, 
the consequences are much milder and the BSM-biased PDFs are statistically undistinguishable from the {\it true} SM PDFs, 
thus effectively disentangling PDF effects from BSM effects.

\end{document}